\documentstyle[12pt]{article}
\catcode`\@=11

\@addtoreset{equation}{section}
\def\theequation{\arabic{section}.\arabic{equation}}

\catcode`\@=11

\def\section{\@startsection{section}{1}{\z@}{3.5ex plus 1ex minus
   .2ex}{2.3ex plus .2ex}{\large\bf}}

\def\eqnarray{\let\@currentlabel=\theequation\refstepcounter{equation}
    \global\@eqnswtrue
    \global\@eqcnt\z@\tabskip\@centering\let\\=\@eqncr
    $$\halign to \displaywidth\bgroup\@eqnsel\hskip\@centering
      $\displaystyle\tabskip\z@{##}$&\global\@eqcnt\@ne
       \hfil${{}##{}}$\hfil
      &\global\@eqcnt\tw@ $\displaystyle\tabskip\z@{##}$\hfil
       \tabskip\@centering&\llap{##}\tabskip\z@\cr}
\def\lefteqn#1{\hbox to 4\arraycolsep{$\displaystyle #1$\hss}}
\def\thesection{\arabic{section}}

\def\appendix{\setcounter{section}{0}
        \def\thesection{Appendix}
        \def\theequation{\Alph{section}.\arabic{equation}}}

\long\def\@makefntext#1{\parindent 0cm\noindent
\hbox to 1em{\hss$^{\@thefnmark}$}#1}
\def\IR{{\hbox{{\rm I}\kern-.2em\hbox{\rm R}}}}
\def\IH{{\hbox{{\rm I}\kern-.2em\hbox{\rm H}}}}
\def\IC{{\ \hbox{{\rm I}\kern-.6em\hbox{\bf C}}}}
\def\IZ{{\hbox{{\rm Z}\kern-.4em\hbox{\rm Z}}}}
\def\rref#1{(\ref{#1})}
\def\comp{{\scriptstyle\circ}}
\newcommand{\beq}{\begin{equation}}
\newcommand{\eeq}{\end{equation}}
\newcommand{\obs}{{\cal O}}
\newcommand{\Obs}{\{{\cal O}\}}

\newcommand{\limh}{\lim_{\hbar\rightarrow0}}
\newcommand{\limht}{\widetilde{\lim_{\hbar\rightarrow0}}}
\newtheorem{Def}{Definition}
\newtheorem{asu}{Assumption}
\newtheorem{lem}{Lemma}
\newtheorem{cond}{Condition}
\newtheorem{theo}{Claim}
\newcommand{\NPB}[1]{{\sl Nucl.~Phys.}~{\bf B#1}}
\newcommand{\Ann}[1]{{\sl Ann.~Phys.}~{\bf #1}}

\newcommand{\CQG}[1]{{\sl Class.~Quant.~Grav.}~{\bf #1}}
\newcommand{\PRD}[1]{{\sl Phys.~Rev.}~{\bf D#1}}

\newcommand{\JMP}[1]{{\sl J.~Math.~Phys.}~{\bf #1}}
\begin{document}
%
%
%
%
\def\citen#1{%
\edef\@tempa{\@ignspaftercomma,#1, \@end, }
\edef\@tempa{\expandafter\@ignendcommas\@tempa\@end}%
\if@filesw \immediate \write \@auxout {\string \citation {\@tempa}}\fi
\@tempcntb\m@ne \let\@h@ld\relax \let\@citea\@empty
\@for \@citeb:=\@tempa\do {\@cmpresscites}%
\@h@ld}
%
\def\@ignspaftercomma#1, {\ifx\@end#1\@empty\else
   #1,\expandafter\@ignspaftercomma\fi}
\def\@ignendcommas,#1,\@end{#1}
%
%
\def\@cmpresscites{%
 \expandafter\let \expandafter\@B@citeB \csname b@\@citeb \endcsname
 \ifx\@B@citeB\relax 
    \@h@ld\@citea\@tempcntb\m@ne{\bf ?}%
    \@warning {Citation `\@citeb ' on page \thepage \space undefined}%
 \else
    \@tempcnta\@tempcntb \advance\@tempcnta\@ne
    \setbox\z@\hbox\bgroup 
    \ifnum\z@<0\@B@citeB \relax
       \egroup \@tempcntb\@B@citeB \relax
       \else \egroup \@tempcntb\m@ne \fi
    \ifnum\@tempcnta=\@tempcntb 
       \ifx\@h@ld\relax 
          \edef \@h@ld{\@citea\@B@citeB}%
       \else 
          \edef\@h@ld{\hbox{--}\penalty\@highpenalty \@B@citeB}%
       \fi
    \else   
       \@h@ld \@citea \@B@citeB \let\@h@ld\relax
 \fi\fi%
 \let\@citea\@citepunct
}
%
\def\@citepunct{,\penalty\@highpenalty\hskip.13em plus.1em minus.1em}%
%
%
\def\@citex[#1]#2{\@cite{\citen{#2}}{#1}}%
%
%
\def\@cite#1#2{\leavevmode\unskip
  \ifnum\lastpenalty=\z@ \penalty\@highpenalty \fi 
  \ [{\multiply\@highpenalty 3 #1
      \if@tempswa,\penalty\@highpenalty\ #2\fi 
    }]\spacefactor\@m}
\let\nocitecount\relax  
%
\begin{titlepage}
\vspace{.5in}
\begin{flushright}
UCD-95-42\\
November 1995\\
\end{flushright}
\vspace{.5in}
\begin{center}
{\Large\bf
 Consistent Evolution with Different Time-Slicings\\[1ex]
 in Quantum Gravity}\\
\vspace{.4in}
{R.~C{\sc osgrove}\\
       {\small\it Department of Physics}\\
       {\small\it University of California}\\
       {\small\it Davis, CA 95616}\\{\small\it USA}}\\
       {\small\it email: cosgrove@bethe.ucdavis.edu}
\end{center}

\vspace{.5in}
\begin{center}
\begin{minipage}{5in}
\begin{center}
{\large\bf Abstract}
\end{center}
{\small
Rovelli's  `` quantum mechanics without time''
motivates an intrinsically
time-slicing independent picture of reduced phase space quantum gravity,
which may be described as  ``quantization after evolution''.
Sufficient criteria for carrying out quantization after evolution are developed
in terms of a general concept of the classical limit of quantum mechanics.
If these criteria are satisfied then it is possible to have consistent
unitary evolution of operators, with respect to an infinite parameter
family of time-slicings (and probably all time-slicings),
with the correct classical limit.  The criteria are particularly
amenable to study in (2+1)-dimensional gravity, where the reduced
phase space is finite dimensional.
}
\end{minipage}
\end{center}
\end{titlepage}

\part{A Time-Slicing Independent Picture of Reduced Phase Space Quantum
Gravity}
\label{p1}

\section{Introduction}
\label{Int}

In reduced phase space quantum gravity a time-slicing is fixed,
and evolution considered with respect to it.
If a different
time-slicing connecting the same initial and final slices is used instead,
it is natural to ask whether the two evolutions will agree.  While in classical
geometrodynamics the answer is yes---after all, the evolution is induced by
a single spacetime metric---in quantum geometrodynamics the answer is in
general no.  Because of operator ordering ambiguities, an operator valued
spacetime metric does not  induce unique operator valued metrics on the
slices of a time-slicing.  A better question is whether evolution along
different time-slicings can be made to agree by judicious choice of the
operator orderings of the Hamiltonians.

Actually, this general problem involves three distinct problems.  In the
terminology of Kucha\v{r} \cite{Kuk} and Isham \cite {Ish1}, these are the
global time problem, the
multiple choice problem, and the functional evolution problem.

The global time problem concerns whether there exists a canonical
transformation
\beq
(g_{ab},K^{ab})\rightarrow (X^{A},P_{A},\phi^{r},p_{r})
\label{a1}
\eeq
 from the intrinsic geometry $g_{ab}$ and extrinsic curvature
$K^{ab}$ of a spacelike hypersurface $\Sigma\hookrightarrow M$, to
``internal spacetime coordinates''
$X^{A}$ which determine the embedding $\Sigma\hookrightarrow M$,
their canonical conjugates $P_{A}$, and the coordinates on the reduced
phase space $(\phi^{r},p_{r})$.  We assume that the canonical
transformation \rref{a1} exists.  Here the term ``reduced phase space'' means
the space obtained by solving the
constraints on the hypersurface $\Sigma\hookrightarrow M$ and
modding out by the transformations generated by the momentum constraints.
Fixing the embedding $\Sigma\hookrightarrow M$ may be interpreted as gauge
fixing the Hamiltonian constraint.

The multiple choice problem concerns the fact that the canonical transformation
\rref{a1} may not be unique.  Fixing the internal coordinates $X^{A}$
determines an embedding $\Sigma\hookrightarrow M$ only after the data
$(\phi^{r},p_{r})$ on $\Sigma$ have been specified and the equations of
geometrodynamics solved.  That is, fixing the internal coordinate functions
$X^{A}$ determines a spacetime metric dependent embedding
$\Sigma\hookrightarrow M$, which we call a slice.
If \rref{a1} is not unique, then given a spacetime metric we can use
internal coordinates $X^{\prime A}$
obtained from a different canonical transformation \rref{a1} to fix the same
embedding.  However, the metric dependence of the embedding may be
different. That is the slice may be different.
It seems possible that using two different canonical
transformations \rref{a1} to fix time-slicings, that is spacetime metric
dependent foliations with spacelike leaves, may result in two different
time-slicings which share the same end slices \cite{Kuk}.  The question of
whether or not evolution of the wave function will agree for the different
time-slicings can be raised here. This is not the problem we are interested
in. We assume that a particular choice of canonical transformation \rref{a1}
has been made.

The problem of functional evolution remains even after the canonical
transformation \rref{a1} has been fixed.  The embedding variables $X^{A}$
can be used to specify various time-slicings.  For each
time-slicing gravity can, in theory, be reduced to a Hamiltonian system by
solving the constraints on each slice and modding out by the transformations
generated by the momentum constraints.  The reduced phase space
$(\phi^{r},p_{r})$ plays the role of an ordinary phase space of an
unconstrained Hamiltonian system, with time a parameter labeling slices.
An example of this reduction has been carried out by Moncrief \cite{Mon}, and
by Hosoya and Nakao \cite{Nak}, for (2+1)-dimensional gravity.  Hence, we
can again arrive at the
situation of having the same end slices connected by many different
time-slicings.  We come now back to our question, can evolution
with respect to all the
different time-slicings be made to agree by judicious choice of the operator
orderings of the Hamiltonians?

Rovelli's quantum mechanics without time \cite{Rov} suggests a way to think
about this problem.  In time reparametrization-invariant systems time
evolution takes the form of a gauge transformation.  In this context,
Rovelli advocates regarding observables not as objects
which take different values at different times (gauges), but
rather as objects whose values are independent of time (gauge).
The traditional
observables are broken up into a sequence of observables, one for each time.
The position of a particle at 1:00 PM is independent of time.
Rovelli proposes regarding such objects as observables.

In this work Rovelli's observables, sometimes called evolving constants of the
motion, are interpreted in terms of covariant canonical quantization
\cite{Ash,Wit1}.  A traditional observable is a function on phase space---the
same function for all times.  The ``value'' of a traditional observable
changes with time because the point in phase space changes.  Rovelli's
observables may be interpreted as functions on the space of solutions of the
classical equations of motion.  Of course a solution does not change
with time; it is the whole path in phase space.  To get evolution the
function on solution space must change with time.  For each traditional
observable, that is each function on phase space, there correspond many
functions on solution space, one for each time.  For example, if the
traditional observable is the position of a particle,
one of the functions on solution
space, when evaluated at a particular solution, yields the position
at 1:00 PM, another yields the position at 2:00 PM etc.

  For a particular traditional observable in gravity, that is a particular
function on the reduced phase space, we obtain a family of functions on
solution space---one for each slice.  Together, the functions on solution
space constitute a consistent evolution system.  Traditionally in quantization,
the function on phase space is raised to operator status.  Here we propose
to lift the whole family of functions on solution space, slice by slice,
to operators, thus obtaining a consistent operator evolution system.  We call
this procedure quantization after evolution because the family of functions
on solution space represent the complete already evolved classical system.
The procedure is
motivated by Carlip's \cite{Car1} (see also \cite{Car2}) comparison of the
Witten \cite{Wit2}, and Moncrief-Hosoya-Nakao \cite{Mon,Nak}
quantizations of (2+1)-gravity.  Kucha\v{r} \cite{Kuk} also mentions this
procedure in his discussion of Rovelli's work.

Quantization after evolution will be defined herein in terms of a general
concept of the classical limit of quantum mechanics \cite{Wer,Ber}.
It gives a Heisenberg
picture of quantum mechanics, in the sense that operators evolve while the
state remains fixed.
However, it is not completely clear that it is equivalent to
evolution by the Heisenberg equations of motion. The main issue concerns
the classical limit
of Hamiltonians, which are not properly defined on individual slices.
To find the
Hamiltonian on a slice, the surrounding time-slicing must be known.
Hence, although quantization after evolution yields completely consistent
evolution of operators defined on slices with the right classical limit,
we are unable to prove that the quantum Hamiltonian has the right
classical limit.  One could claim that this is not enough to show that the
functional evolution problem may be solved, because it does not actually
show that the Hamiltonians may be ``operator ordered'' (implying that they
have the right classical limit) to give consistent evolution.
The issue of whether or not Hamiltonians should be considered as observables
in geometrodynamics arises.

After defining quantization after evolution, the problem of constructing a
quantization after evolution will be formulated as the problem of finding a
lifting of canonical transformations to unitary transformations.  This
formulation will be used to show that if certain criteria are satisfied,
then for manifolds of the form $M=\Sigma\times\IR^{+}$ quantization after
evolution may be
implemented on an infinite parameter family of time-slicings, and probably on
all time-slicings.  The main criterion which must be satisfied
is that it must be locally possible to continuously lift real valued
functions on phase space to hermitian operators with the right classical
limit.  The results of this paper mean that if the criteria are
satisfied, then consistent evolution of operators defined on slices
with the right classical limit may
be achieved for at least an infinite parameter family of time-slicings.
In addition it is shown that any particular time-slicing, with evolution
defined by any operator ordering of the associated Hamiltonian, may be
included in such a consistent infinite parameter family.

The article is organized into three broad parts.  Part I gives a conceptual
introduction to the intrinsically time-slicing independent picture of reduced
phase space quantum gravity mentioned above.  Part II makes the ideas of
part I precise by defining quantization after evolution and formulating the
question of existence.  Part III discusses the question of existence, deriving
the result on consistent evolution.


\section{Motivation}
\label{Mot}

Consider the solution of a classical evolution problem on a two dimensional
phase space by means of Hamilton-Jacobi theory. Find a (time-dependent)
canonical transformation
 from the variables $(q,p)$ to variables $(\alpha,\beta)$, for
which the Hamiltonian is zero.  Starting with initial data
$(q(t_{0}),p(t_{0}))$, implement the canonical transformation at time
$t_{0}$ to find
\beq
\left(\alpha(t_{0})=\alpha_{*t_{0}}(q(t_{0}),p(t_{0})),\
\beta(t_{0})=\beta_{*t_{0}}(q(t_{0}),p(t_{0}))\right).
\label{b1}
\eeq
Evolve $(\alpha,\beta)$ (they don't change since H=0).
Then implement the inverse canonical
transformation at time t to find
\beq
\left(q(t)=q_{*t}(\alpha(t),\beta(t)),
\ p(t)=p_{*t}(\alpha(t),\beta(t))\right).
\label{b2}
\eeq

A natural question seems to be: Can we use the above method to solve the
evolution problem in Heisenberg picture quantum mechanics?  To this end,
quantize the classical mechanical system in the variables $(\alpha,\ \beta)$:
$\alpha\mapsto\hat{\alpha}$, $\beta\mapsto\hat{\beta}$,
$0=H(t)\mapsto\hat{H}(t)$.  If ever there were a natural operator ordering for
$\hat{H}(t)$ there is here, $\hat{H}(t)=0$. So $(\hat{\alpha},\ \hat{\beta})$
are constant in time, just like their classical counterparts.

Therefore, starting with the initial operators $(\hat{q}(t_{0}),\ \hat{p}
(t_{0}))$, use the canonical transformation \rref{b1}, elevated to an
operator equation, to find
\beq
\left(\hat{\alpha}(t_{0})=\alpha_{*t_{0}}(\hat{q}(t_{0}),\hat{p}(t_{0})),\
\hat{\beta}(t_{0})=\beta_{*t_{0}}(\hat{q}(t_{0}),\hat{p}(t_{0}))\right).
\label{b3}
\eeq
Evolve $(\hat{\alpha},\hat{\beta})$ (they don't change since $\hat{H}=0$).
Then implement the inverse canonical transformation at time t \rref{b2},
elevated to an operator equation,  to find
\beq
\left(\hat{q}(t)=q_{*t}(\hat{\alpha}(t),\hat{\beta}(t)),\
\hat{p}(t)=p_{*t}(\hat{\alpha}(t),\hat{\beta}(t))\right).
\label{b4}
\eeq

Unfortunately, equations \rref{b3} and \rref{b4} are not uniquely defined
by equations \rref{b1} and \rref{b2}.  A choice of operator ordering must
be made.   However, an
interesting question has been spawned. To what extent is the operator
ordering ambiguity associated with \rref{b3} and \rref{b4} equivalent to
the operator ordering ambiguity associated with making $H(q,p,t)$ into an
operator $\hat{H}(t)$ and solving the Heisenberg equations of motion to find
$(\hat{q}(t),\ \hat{p}(t))$ in terms of $(\hat{q}(t_{0}),\ \hat{p}(t_{0})
)$?  (Clearly they are not equivalent unless the operator orderings
in \rref{b3} and \rref{b4} are restricted to give unitary evolution.)

The above discussion motivates a picture of evolution in quantum mechanics
which I call {\em quantization after evolution}.  In the following two
sections this discussion is generalized and a rough preliminary definition of
quantization after evolution given.

\section{Already Evolved Classical System}
\label{Alr}

With the right notation and definitions we can take the evolution out of
evolution.  The symbols $p$ and $q$ are most properly regarded as representing
coordinate functions on phase space $PS$, that is mappings
\begin{eqnarray*}
q:PS&\rightarrow& \IR,\\
p:PS&\rightarrow& \IR.
\end{eqnarray*}
What is meant by $q(t)$ and $p(t)$ in the previous example?  We like to
think of evolution as a path in the phase space $\xi:\IR\rightarrow PS$.
The symbols $q(t)$ and $p(t)$ are short for $q(\xi(t))$ and $p(\xi(t))$,
which are real numbers (for each fixed t).
Similarly $\alpha(t)$ and $\beta(t)$ must be real
numbers; but how should we interpret them?

The solution space $SS$ is the space of parametrized (by time) paths in
phase space which satisfy the equations of motion.  (Here we are speaking
of an unconstrained Hamiltonian system.)  That is, an element $\xi$ of
$SS$ is a mapping
\beq
\xi:\IR\rightarrow PS.
\label{sel}
\eeq
Given a solution $\xi\in SS$,
real numbers $\alpha(t)$ and $\beta(t)$ are determined which are {\em
independent of time}.  Hence $\alpha(t)$ and $\beta(t)$ are really just
bad notation for $\alpha(\xi)$ and $\beta(\xi)$.  That is, the symbols
$\alpha$
and $\beta$ are most properly regarded as representing coordinate functions
on solution space, that is mappings
\begin{eqnarray*}
\alpha:SS&\rightarrow&\IR,\\
\beta:SS&\rightarrow&\IR.
\end{eqnarray*}

Now that we have straightened things out a bit, define the the following pair
of functions on solution space by substituting $\alpha$ for $\alpha(t)$
and $\beta$ for $\beta(t)$ in equations \rref{b2}:
\beq
\left(q_{t}=q_{*t}(\alpha,\beta),
\ p_{t}=p_{*t}(\alpha,\beta)\right).
\eeq
That is:
\begin{eqnarray}
q_{t}:SS&\rightarrow&\IR\nonumber\\
\xi&\mapsto&q_{*t}(\alpha(\xi),\beta(\xi))\nonumber\\
p_{t}:SS&\rightarrow&\IR\nonumber\\
\xi&\mapsto&p_{*t}(\alpha(\xi),\beta(\xi)).
\end{eqnarray}
For the mapping $q$, $q_{t}(\xi)$ is the image of the point in phase space
$\xi(t)$:
\beq
q_{t}(\xi)=q(\xi(t)).
\label{sso}
\eeq
The mapping $p$ and $p_{t}(\xi)$ have the same relation.
In sloppy language we may say:  $q_{t}(\xi)$
is the ``value of $q$ at time $t$'' when the solution is $\xi$.

Let us now generalize the concept of the functions on solution space
$q_{t}$ and $p_{t}$ to any phase space and any observable.  We still only
deal in unconstrained Hamiltonian systems.  A function on phase space
\beq
\obs:PS\rightarrow\IR
\eeq
will be called a phase space observable.  $q$ and $p$ are special examples.
The pair $(q_{t},t)$ will be called the solution space observable
corresponding to the phase space observable $q$ at time $t$, in the example.
Generally, we define the solution space observable corresponding to the
phase space observable $\obs$ at time $t$ to be the pair $(\obs_{t},t)$
where $\obs_{t}$ is defined by analogy with \rref{sso}.  Given the mapping
$\obs:PS \rightarrow \IR$ and the time $t$, $\obs_{t}$ is the mapping
$\obs_{t}:SS\rightarrow\IR$ defined by
\beq
\obs_{t}(\xi)=\obs(\xi(t)),
\label{def}
\eeq
where $\xi(t)$ is the point in phase space corresponding to the
solution $\xi$ and the time $t$.  The mappings $\obs_{t}$ are essentially
equivalent to Rovelli's evolving constants of the motion \cite{Rov}.
However, the definition given here is more closely related to the
covariant canonical quantization point of view \cite{Ash,Wit1}.

\begin{picture}(400,275)(-110,-70)
\newsavebox{\hor}

\savebox{\hor}(0,0)[bl]{\thicklines{\line(1,0){125}}}

\multiput(0,0)(0,50){5}{\usebox{\hor}}

\put(62.5,4){\makebox(0,0)[b]{$t_{4}$}}
\put(62.5,54){\makebox(0,0)[b]{$t_{3}$}}
\put(62.5,104){\makebox(0,0)[b]{$t_{2}$}}
\put(62.5,154){\makebox(0,0)[b]{$t_{1}$}}
\put(62.5,204){\makebox(0,0)[b]{$t_{0}$}}

\put(135,5){\makebox(0,0)[bl]{$q_{\ast t_{4}}(\alpha,\beta)$}}
\put(135,0){\makebox(0,0)[tl]{$p_{\ast t_{4}}(\alpha,\beta)$}}
\put(135,55){\makebox(0,0)[bl]{$q_{\ast t_{3}}(\alpha,\beta)$}}
\put(135,50){\makebox(0,0)[tl]{$p_{\ast t_{3}}(\alpha,\beta)$}}
\put(135,105){\makebox(0,0)[bl]{$q_{\ast t_{2}}(\alpha,\beta)$}}
\put(135,100){\makebox(0,0)[tl]{$p_{\ast t_{2}}(\alpha,\beta)$}}
\put(135,155){\makebox(0,0)[bl]{$q_{\ast t_{1}}(\alpha,\beta)$}}
\put(135,150){\makebox(0,0)[tl]{$p_{\ast t_{1}}(\alpha,\beta)$}}
\put(135,205){\makebox(0,0)[bl]{$q_{\ast t_{0}}(\alpha,\beta)$}}
\put(135,200){\makebox(0,0)[tl]{$p_{\ast t_{0}}(\alpha,\beta)$}}
\put(-40,-40){Figure 1\ \ \ \ The Already Evolved Classical System}
\sbox{\hor}{}
\end{picture}

The reason for calling the pair $(\obs_{t},t)$ the solution space
observable instead of $\obs_{t}$ is that $(\obs_{t},t)$ is distinguished
 from $(\obs_{t^{\prime}},t^{\prime})$ when $t\neq t^{\prime}$ even if
$\obs_{t}=\obs_{t^{\prime}}$. Now we define the already evolved classical
system $(AES)$ as the set of all such pairs:
\begin{eqnarray}
AES=&\{&(\obs_{t},t)\ |\ t\in\IR,\ \obs_{t}\in SS^{*}\},\nonumber\\
\ &(&SS^{*}\ denotes\ the\ dual\ space\ of\ SS).
\label{c9}
\end{eqnarray}
The already evolved classical system is represented schematically in
Figure 1.  For brevity, Figure 1 represents the phase space observables
$q$ and $p$ from the motivating example.  Note, however, that the
definitions of $\obs_{t}$ and the $AES$ do not depend at all on this
example.

The name ``already evolved classical system'' is meant to be descriptive.
Input a solution, and out pops the whole already evolved system.  The $AES$
does not evolve; however it has a causal structure.
If we let $\{\obs\}$ be the set of phase space observables, then
\rref{def} gives a natural bijection mapping $\{\obs\}\times\IR$
to the $AES$:
\begin{eqnarray}
B:\{\obs\}\times\IR&\rightarrow& AES,\nonumber\\
B(\obs,t)&=&(\obs_{t},t).
\label{e1}
\end{eqnarray}
$B$ induces a Poisson bracket on the space of functions on solution space
$\{\obs_{t}\}$ from the Poisson bracket on the space of functions on
phase space $\{\obs\}$.  Specifically, we fix $t=t_{0}$ and define
\beq
\left(\{\obs_{1t_{0}},\obs_{2t_{0}}\}_{pb},t_{0}\right)=
B\left(\{\obs_{1},\obs_{2}\}_{pb},t_{0}\right).
\eeq
The definition is independent of $t_{0}$ because time evolution is a
canonical transformation.  Of course we can also look at this as the solution
space acquiring a symplectic structure from the phase space.

\begin{picture}(400,275)(-110,-70)

\savebox{\hor}(0,0)[bl]{\thicklines{\line(1,0){125}}}

\multiput(0,0)(0,50){5}{\usebox{\hor}}

\put(62.5,4){\makebox(0,0)[b]{$t_{4}$}}
\put(62.5,54){\makebox(0,0)[b]{$t_{3}$}}
\put(62.5,104){\makebox(0,0)[b]{$t_{2}$}}
\put(62.5,154){\makebox(0,0)[b]{$t_{1}$}}
\put(62.5,204){\makebox(0,0)[b]{$t_{0}$}}

\put(135,5){\makebox(0,0)[bl]{$q_{\ast t_{4}}(\hat{\alpha},\hat{\beta})$}}
\put(135,0){\makebox(0,0)[tl]{$p_{\ast t_{4}}(\hat{\alpha},\hat{\beta})$}}
\put(135,55){\makebox(0,0)[bl]{$q_{\ast t_{3}}(\hat{\alpha},\hat{\beta})$}}
\put(135,50){\makebox(0,0)[tl]{$p_{\ast t_{3}}(\hat{\alpha},\hat{\beta})$}}
\put(135,105){\makebox(0,0)[bl]{$q_{\ast t_{2}}(\hat{\alpha},\hat{\beta})$}}
\put(135,100){\makebox(0,0)[tl]{$p_{\ast t_{2}}(\hat{\alpha},\hat{\beta})$}}
\put(135,155){\makebox(0,0)[bl]{$q_{\ast t_{1}}(\hat{\alpha},\hat{\beta})$}}
\put(135,150){\makebox(0,0)[tl]{$p_{\ast t_{1}}(\hat{\alpha},\hat{\beta})$}}
\put(135,205){\makebox(0,0)[bl]{$q_{\ast t_{0}}(\hat{\alpha},\hat{\beta})$}}
\put(135,200){\makebox(0,0)[tl]{$p_{\ast t_{0}}(\hat{\alpha},\hat{\beta})$}}
\put(-40,-40){Figure 2\ \ \ \ Quantization After Evolution}
\sbox{\hor}{}
\end{picture}

An explicit construction of solution space observables for the constant mean
curvature time-slicing of (2+1)-gravity
is given by Carlip \cite{Car1}.  In this construction, Witten's holonomy
variables \cite{Wit2} can be viewed as coordinates on the solution space.
Equivalently, Witten's
holonomy variables can be viewed as the Hamilton-Jacobi variables.

\section{Quantization After Evolution I}
\label{Qua1}

Quantization after evolution is quantizing the already evolved classical
system.  For simplicity, let us return to our two dimensional example.
One way to perform quantization after evolution is to quantize the solution
space in the
canonical variables
$(\alpha,\beta)$:
\begin{eqnarray}
(\alpha,\beta)&\mapsto&(\hat{\alpha},\hat{\beta}),\nonumber\\
\{\alpha,\ \beta\}_{pb}&\mapsto&\frac{1}{i\hbar}[\hat{\alpha},\ \hat{\beta}],
\label{c2}
\end{eqnarray}
then substitute the operators $(\hat{\alpha},\hat{\beta})$ for
$(\alpha,\beta)$ in the functions on solution space
which comprise the $AES$.
This is represented schematically in Figure 2, which is Figure 1 with hats
added. In doing
this we should choose operator orderings which respect the canonical
structure of $(q,p)$, and the causal structure of the $AES$.  Specifically
we require
\begin{eqnarray}
\frac{1}{i\hbar}[\hat{q}_{t}&,&\ \hat{p}_{t}]=1\ \ and,\nonumber\\
\hat{q}_{t^{\prime}}=U\hat{q}_{t}U^{\dag}&,&
\ \hat{p}_{t^{\prime}}=U\hat{p}_{t}U^{\dag},
\ where\ UU^{\dag}=1.
\label{c3}
\end{eqnarray}

Due to results such as Van Hove's theorem \cite{obs,geo}, the success of
this venture will
depend on the parametrization $(\alpha,\beta)$ of the solution space.  Because
of this problem, a slightly different and  more general definition
of quantization after evolution will be given in Section~\ref{Qua}, and used
for
the analysis portion of this paper.

An explicit construction of a quantization after evolution has been carried
out by Carlip \cite{Car1} for the example mentioned above, in the context of
comparing Moncrief's and Witten's quantizations of (2+1)-gravity.
The relation here is that Witten's quantization can be viewed as a
quantization of solution space
$((\alpha,\beta)\mapsto(\hat{\alpha},\hat{\beta}))$,
while Moncrief's quantization can be viewed as a quantization of the
(reduced) phase space and subsequent evolution under the
Heisenberg equations of motion.  Although
he does not name his construction, Carlip essentially constructs a
quantization after evolution, with Witten's holonomy variables parametrizing
the solution space, and finds an operator ordering of Moncrief's
Hamiltonian such that the Heisenberg equation evolution agrees with the
quantization after evolution.

This example further motivates the question: is quantization after evolution
equivalent, in general, to quantizing the phase space and evolving
by means of the Heisenberg equations of motion.  This question will be
considered after more formal definitions have been given
(Sections~\ref{Hei}--~\ref{Com}).
The next section begins discussion of systems which don't have a
preferred time parameter.

\section{Systems Without Preferred Time Parameter}
\label{Sys}

For the purposes of this paper,
the most important feature of quantization after evolution
is that it applies to systems without a preferred time
parameter and leads to intrinsically consistent evolution.  So far we
have studied Hamiltonian systems.  For a Hamiltonian system the space of
time is $\IR$.  A solution $\xi$ in $SS$ is a mapping
\beq
\label{eham}
\xi : \IR\rightarrow PS.
\eeq
Gravity can be made to fit this mold by choosing a time-slicing. On each
slice in a one parameter $(\IR)$ family of slices a spatial metric and
extrinsic curvature tensor (point in $PS$) are induced by a spacetime
metric (point in $SS$).  (This statement is meant for illustration only;
the term ``slice'' requires explanation.
A more careful discussion involving the reduced phase space is given in
the next section.)  However this reduction is unnatural. Which time-slicing
should we choose?  More naturally, gravity fits a generalization of this
mold. On each slice in the space of {\em all} slices $(S)$ a spatial metric
and extrinsic curvature tensor (point in $PS$) are induced by a spacetime
metric (point in $SS$).  That is for gravity, a solution $\xi$ in $SS$
is a mapping
\beq
\xi : S\rightarrow PS.
\label{egen}
\eeq

Much of what follows will be valid either in the situation \rref{eham} or
in the situation \rref{egen}.  In fact it will be valid for any ``system
with global time $\Gamma$'':
\begin{Def}
\label{dgen}
A system with global time $\Gamma$ is a triple
$(PS,\ SS,\ \Gamma)$ where $PS$, $SS$, and $\Gamma$ are topological spaces
with differential structure such that:
\begin{enumerate}
\item
An element $\xi$ of $SS$ is a differentiable mapping
$$\xi :\Gamma\rightarrow PS.$$
\item
$PS$ is a symplectic space. For all $\tau$ and $\tau^{\prime}$ in
$\Gamma$, $\xi(\tau)$ and $\xi(\tau^{\prime})$ are related by a canonical
transformation $u(\tau,\tau^{\prime})$, which does not depend on $\xi$.
\item
For every $\tau$ in $\Gamma$ and $\eta$ in $PS$ there exists $\xi$ in
$SS$ with $\xi(\tau)=\eta$.
\end{enumerate}
\end{Def}
If $\Gamma$ is not $\IR$ we will also call this system a ``system without
preferred time parameter''.  Of course we will call $PS$ the phase space,
$SS$ the solution space, and $\Gamma$ the space of global time.  This
definition is introduced so that we may carry on our discussion in general
terms, and then when appropriate set $\Gamma$ equal to $\IR$,
$S$, or a subspace of $S$.  It will also be used
in the next section to state an important assumption regarding
the nature of time in gravity.

It is useful to describe the consistent evolution problem in the general
setting of a system with global time $\Gamma$.
A Hamiltonian system may be obtained from a system with global time
$\Gamma$ by choosing a ``preferred time'', that is by choosing a
differentiable embedding
$\IR\hookrightarrow\Gamma$.  For gravity, a choice of time-slicing is a
special case of a choice of preferred time.
Of course a Hamiltonian system can be quantized in the usual Heisenberg
picture.  If this is done for more then one choice of preferred time,
specifically for two embeddings $\IR\stackrel{e_{1}}{\hookrightarrow}\Gamma$
and $\IR\stackrel{e_{2}}{\hookrightarrow}\Gamma$ which coincide at two
points, the operator evolution between the two points will not agree in
general.
The question naturally arises:  can we operator order the two Hamiltonians
so that the operator evolution agrees?  More generally:  can we operator
order the infinitude of Hamiltonians corresponding to all possible
embeddings $\IR\hookrightarrow\Gamma$ so that the evolution is completely
consistent?  This question will be referred to as the question of achieving
consistent evolution in the Heisenberg picture for a quantum theory
without preferred time parameter.  Of course our emphasis is on gravity.
It is simply convenient to speak in general terms.

It is  interesting to note that most of the systems studied in physics have
(probable) interpretations as systems of global time $\Gamma$ for
some $\Gamma$.  For example, special relativistic systems have a (probable)
interpretation as systems of global time $\Gamma$ with $\Gamma$ the space
of flat spacelike planes in Minkowski space.  In an upcoming paper it will
be shown explicitly that the relativistic free particle system in
(1+1)-dimensions is a system with global time $\IR\times\IR$.  The system
is found to have two Hamiltonians associated with two time variables which
parametrize the straight spacelike lines.  A consistent quantization (in the
sense just described) is carried out explicitly and the result appears to be
a viable single particle interpretation of relativistic quantum mechanics in
(1+1)-dimensions.

The definitions given in Section~\ref{Alr} are applicable to
any system
with global time $\Gamma$.  Simply substitute $\Gamma$ for $\IR$ and
the words ``global time'' for ``time''.  For definiteness, let us write
them here as formal definitions.
\begin{Def}
\label{evo}
Given the mapping
$\obs:PS \rightarrow \IR$ and the global time $\tau\in\Gamma$,
$\obs_{\tau}$ is the mapping
$\obs_{\tau}:SS\rightarrow\IR$ defined by
\beq
\obs_{\tau}(\xi)=\obs(\xi(\tau)),\nonumber
\eeq
where $\xi(\tau)$ is the point in phase space corresponding to the
solution $\xi$ and the global time $\tau$.
\end{Def}

\begin{picture}(300,275)(-10,-70)

\newsavebox{\hill}
\newsavebox{\mount}

\savebox{\hor}(0,0)[bl]{\thicklines{\line(1,0){125}}}

\multiput(0,0)(0,50){5}{\usebox{\hor}}

\put(62.5,4){\makebox(0,0)[b]{$t_{4}$}}
\put(62.5,54){\makebox(0,0)[b]{$t_{3}$}}
\put(62.5,104){\makebox(0,0)[b]{$t_{2}$}}
\put(62.5,154){\makebox(0,0)[b]{$t_{1}$}}
\put(62.5,204){\makebox(0,0)[b]{$t_{0}$}}
\put(287.5,4){\makebox(0,0)[b]{$t_{4}$}}
\put(287.5,55){\makebox(0,0)[b]{$t_{3}^{\prime}$}}
\put(287.5,105){\makebox(0,0)[b]{$t_{2}^{\prime}$}}
\put(287.5,155){\makebox(0,0)[b]{$t_{1}^{\prime}$}}
\put(287.5,204){\makebox(0,0)[b]{$t_{0}$}}

\put(135,5){\makebox(0,0)[bl]{$q_{\ast t_{4}}(\alpha,\beta)$}}
\put(135,0){\makebox(0,0)[tl]{$p_{\ast t_{4}}(\alpha,\beta)$}}
\put(135,55){\makebox(0,0)[bl]{$q_{\ast t_{3}}(\alpha,\beta)$}}
\put(135,50){\makebox(0,0)[tl]{$p_{\ast t_{3}}(\alpha,\beta)$}}
\put(135,105){\makebox(0,0)[bl]{$q_{\ast t_{2}}(\alpha,\beta)$}}
\put(135,100){\makebox(0,0)[tl]{$p_{\ast t_{2}}(\alpha,\beta)$}}
\put(135,155){\makebox(0,0)[bl]{$q_{\ast t_{1}}(\alpha,\beta)$}}
\put(135,150){\makebox(0,0)[tl]{$p_{\ast t_{1}}(\alpha,\beta)$}}
\put(135,205){\makebox(0,0)[bl]{$q_{\ast t_{0}}(\alpha,\beta)$}}
\put(135,200){\makebox(0,0)[tl]{$p_{\ast t_{0}}(\alpha,\beta)$}}

\put(360,5){\makebox(0,0)[bl]{$q_{\ast t_{4}}(\alpha,\beta)$}}
\put(360,0){\makebox(0,0)[tl]{$p_{\ast t_{4}}(\alpha,\beta)$}}
\put(360,55){\makebox(0,0)[bl]{$q_{\ast t_{3}^{\prime}}(\alpha,\beta)$}}
\put(360,50){\makebox(0,0)[tl]{$p_{\ast t_{3}^{\prime}}(\alpha,\beta)$}}
\put(360,105){\makebox(0,0)[bl]{$q_{\ast t_{2}^{\prime}}(\alpha,\beta)$}}
\put(360,100){\makebox(0,0)[tl]{$p_{\ast t_{2}^{\prime}}(\alpha,\beta)$}}
\put(360,155){\makebox(0,0)[bl]{$q_{\ast t_{1}^{\prime}}(\alpha,\beta)$}}
\put(360,150){\makebox(0,0)[tl]{$p_{\ast t_{1}^{\prime}}(\alpha,\beta)$}}
\put(360,205){\makebox(0,0)[bl]{$q_{\ast t_{0}}(\alpha,\beta)$}}
\put(360,200){\makebox(0,0)[tl]{$p_{\ast t_{0}}(\alpha,\beta)$}}

\put(225,0){\usebox{\hor}}
\put(225,200){\usebox{\hor}}
\savebox{\hill}(0,0){\thicklines{\line(6,-1){125}}}
\savebox{\mount}(0,0){\thicklines{\line(3,-1){125}}}
\put(287.5,50){\usebox{\hill}}
\put(287.5,150){\usebox{\hill}}
\put(287.5,100){\usebox{\mount}}
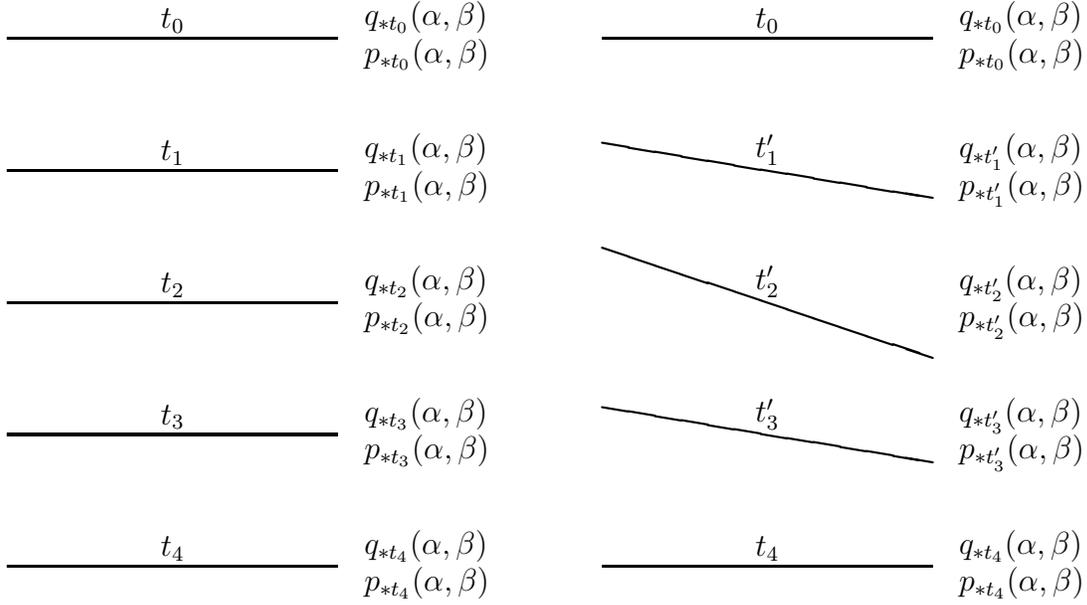
\put(-10,-40){Figure 3\ \ \ \ An already evolved classical system without
preferred time parameter}
\sbox{\hor}{}
\sbox{\hill}{}
\sbox{\mount}{}
\end{picture}

\begin{Def}
\label{Aes}
\begin{eqnarray*}
AES=&\{&(\obs_{\tau},\tau)\ |\ \tau\in\Gamma,\ \obs_{\tau}\in SS^{*}\},\\
\ &(&SS^{*}\ denotes\ the\ dual\ space\ of\ SS).
\end{eqnarray*}
\end{Def}
The bijection $B$ becomes
\begin{eqnarray}
B:\{\obs\}\times\Gamma&\rightarrow& AES,\nonumber\\
B(\obs,\tau)&=&(\obs_{\tau},\tau).
\label{bij}
\end{eqnarray}
It still induces Poisson brackets on the space of functions on solution space.

Quantization after evolution is also applicable to any system
with global time $\Gamma$.
An already evolved classical system without preferred time is represented
schematically in Figure 3, which should be compared with Figure 1.
Quantization
after evolution is shown schematically in Figure 4, which should be compared
with Figure 2,  and proceeds precisely as in Section~\ref{Qua1} except that
$t$ and $t^{\prime}$ in equation \rref{c3} should be replaced by $\tau$ and
$\tau^{\prime}$, which are elements of $\Gamma$.

Of course, the above doesn't prove that it is possible to find ``operator
orderings'' which satisfy the conditions \rref{c3}, that is to
carry out quantization after evolution.  But if we can, the
resulting evolution is intrinsically consistent.
Classical functions on solution space are assigned operator status on an
instant by instant basis (instant of global time of course).

\begin{picture}(300,275)(-10,-70)

\savebox{\hor}(0,0)[bl]{\thicklines{\line(1,0){125}}}

\multiput(0,0)(0,50){5}{\usebox{\hor}}

\put(62.5,4){\makebox(0,0)[b]{$t_{4}$}}
\put(62.5,54){\makebox(0,0)[b]{$t_{3}$}}
\put(62.5,104){\makebox(0,0)[b]{$t_{2}$}}
\put(62.5,154){\makebox(0,0)[b]{$t_{1}$}}
\put(62.5,204){\makebox(0,0)[b]{$t_{0}$}}
\put(287.5,4){\makebox(0,0)[b]{$t_{4}$}}
\put(287.5,55){\makebox(0,0)[b]{$t_{3}^{\prime}$}}
\put(287.5,105){\makebox(0,0)[b]{$t_{2}^{\prime}$}}
\put(287.5,155){\makebox(0,0)[b]{$t_{1}^{\prime}$}}
\put(287.5,204){\makebox(0,0)[b]{$t_{0}$}}

\put(135,5){\makebox(0,0)[bl]{$q_{\ast t_{4}}(\hat{\alpha},\hat{\beta})$}}
\put(135,0){\makebox(0,0)[tl]{$p_{\ast t_{4}}(\hat{\alpha},\hat{\beta})$}}
\put(135,55){\makebox(0,0)[bl]{$q_{\ast t_{3}}(\hat{\alpha},\hat{\beta})$}}
\put(135,50){\makebox(0,0)[tl]{$p_{\ast t_{3}}(\hat{\alpha},\hat{\beta})$}}
\put(135,105){\makebox(0,0)[bl]{$q_{\ast t_{2}}(\hat{\alpha},\hat{\beta})$}}
\put(135,100){\makebox(0,0)[tl]{$p_{\ast t_{2}}(\hat{\alpha},\hat{\beta})$}}
\put(135,155){\makebox(0,0)[bl]{$q_{\ast t_{1}}(\hat{\alpha},\hat{\beta})$}}
\put(135,150){\makebox(0,0)[tl]{$p_{\ast t_{1}}(\hat{\alpha},\hat{\beta})$}}
\put(135,205){\makebox(0,0)[bl]{$q_{\ast t_{0}}(\hat{\alpha},\hat{\beta})$}}
\put(135,200){\makebox(0,0)[tl]{$p_{\ast t_{0}}(\hat{\alpha},\hat{\beta})$}}

\put(360,5){\makebox(0,0)[bl]{$q_{\ast t_{4}}(\hat{\alpha},\hat{\beta})$}}
\put(360,0){\makebox(0,0)[tl]{$p_{\ast t_{4}}(\hat{\alpha},\hat{\beta})$}}
\put(360,55){\makebox(0,0)[bl]{$q_{\ast t_{3}^{\prime}}(\hat{\alpha},
\hat{\beta})$}}
\put(360,50){\makebox(0,0)[tl]{$p_{\ast t_{3}^{\prime}}(\hat{\alpha},
\hat{\beta})$}}
\put(360,105){\makebox(0,0)[bl]{$q_{\ast t_{2}^{\prime}}(\hat{\alpha},
\hat{\beta})$}}
\put(360,100){\makebox(0,0)[tl]{$p_{\ast t_{2}^{\prime}}(\hat{\alpha},
\hat{\beta})$}}
\put(360,155){\makebox(0,0)[bl]{$q_{\ast t_{1}^{\prime}}(\hat{\alpha},
\hat{\beta})$}}
\put(360,150){\makebox(0,0)[tl]{$p_{\ast t_{1}^{\prime}}(\hat{\alpha},
\hat{\beta})$}}
\put(360,205){\makebox(0,0)[bl]{$q_{\ast t_{0}}(\hat{\alpha},\hat{\beta})$}}
\put(360,200){\makebox(0,0)[tl]{$p_{\ast t_{0}}(\hat{\alpha},\hat{\beta})$}}

\put(225,0){\usebox{\hor}}
\put(225,200){\usebox{\hor}}
\savebox{\hill}(0,0){\thicklines{\line(6,-1){125}}}
\savebox{\mount}(0,0){\thicklines{\line(3,-1){125}}}
\put(287.5,50){\usebox{\hill}}
\put(287.5,150){\usebox{\hill}}
\put(287.5,100){\usebox{\mount}}
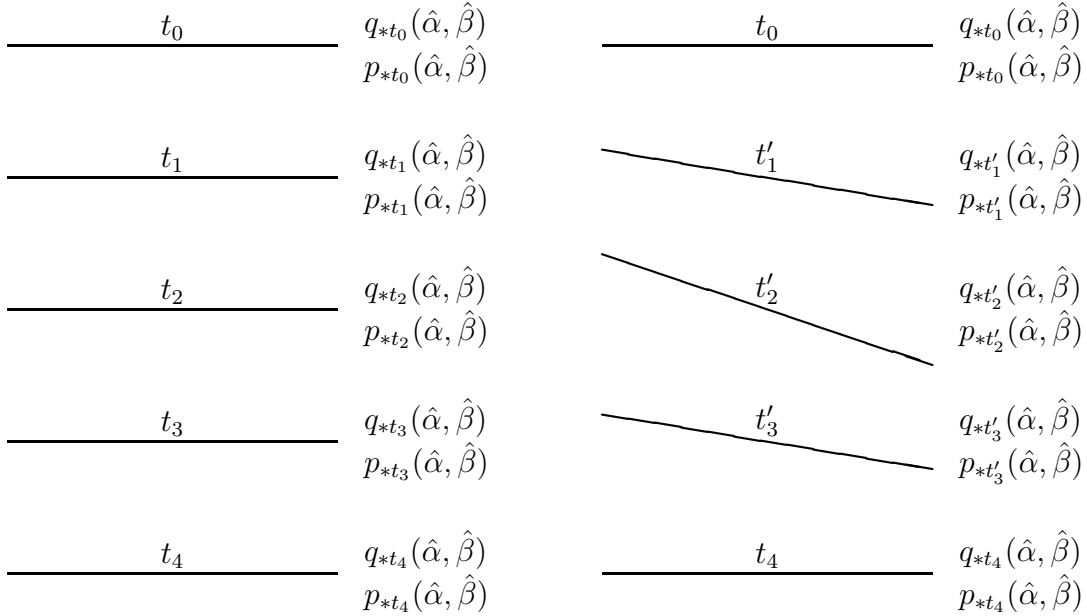
\put(-10,-40){Figure 4\ \ \ \ Quantization after evolution of a system
without preferred time parameter}
\sbox{\hor}{}
\sbox{\hill}{}
\sbox{\mount}{}
\end{picture}

\section{Gravity as a System Without Preferred Time Parameter}
\label{Gra}

Missing from the previous section is a clear description of what is meant
by a ``slice''.  This is a non-trivial question involving the
global time and the multiple choice problems.  In this section
we will first motivate and then state an assumption concerning the nature
of time in gravity. Along the way, the meaning of the term ``slice'' will
be explained.

The definition of the ``reduced phase space'' is the constraint surface modulo
the transformations generated by the first class constraints.  In many
cases finding the reduced phase space reduces a theory to an unconstrained
Hamiltonian system with the reduced phase space playing the role of an
ordinary phase space.  In gravity, however, the symmetries are the
diffeomorphisms of the spacetime manifold.  After finding the {\em fully}
reduced phase space, no equations of motion remain (at least if the spacetime
is spatially closed).  Evolution amounts
to a spacetime diffeomorphism, and hence has been modded out of the fully
reduced phase space.  The fully reduced phase space is thus {\em naturally}
isomorphic to the solution space.\footnote{When equations of motion are
present, the reduced phase space is isomorphic to the solution space, but
not naturally isomorphic.  A different isomorphism is obtained for each time
$t_{0}$ by mapping a point in phase space to the solution for which it is
the initial data at time $t_{0}$.}  The complete reduction has been
carried out for (2+1)-dimensional gravity by Witten \cite{Wit2}.

If an unconstrained Hamiltonian system with phase space coordinates
$(q,p)$ and Lagrangian
$$L=\int (p\frac{dq}{dT}-H)dT$$
is parametrized, then it acquires an equivalent
description as a constrained system with phase space coordinates
$(q,p,T,P_{T},K,P_{K})$ and Lagrangian
$$L^{\prime}=\int (p\frac{dq}{dt}+P_{T}\frac{dT}{dt}-(H+P_{T})K)dt.$$
If the fully reduced phase space is found for this constrained system, the
result is not the original phase space $(q,p)$.  The constraint $H+P_{T}=0$
generates evolution of $q$ and $p$, so that the reduced phase space is
{\em naturally} isomorphic to the solution space just as for gravity.
To regain the original Hamiltonian system, what must be done is to fix
the gauge
$$\frac{dT}{dt}=K=1,$$
not mod out by the transformations generated by the constraint.

In gravity, the analogous gauge condition is the fixing of a time-slicing.
However, after a time-slicing has been fixed, a constrained system still
remains. (The gauge has not been completely fixed.)  If at this point the
reduced phase space is found, the result is an unconstrained Hamiltonian
system with time a parameter labeling slices.  This ``partially reduced phase
space'' is to be distinguished from the fully reduced phase space
mentioned earlier.  This (partial) reduction has
been carried out for the constant mean curvature time-slicing (York time)
in (2+1)-dimensional gravity by Moncrief \cite{Mon}, and by Hosoya and
Nakao \cite{Nak}.

The global time problem concerns whether this reduction can be carried out
in general, for any foliation of the manifold with spacelike leaves.
The general reduction amounts to finding a canonical transformation
\beq
(g_{ab},\ K^{ab})\rightarrow (X^{A},\ P_{A},\ \phi^{r},\ p_{r})
\label{z1}
\eeq
 from the intrinsic geometry $g_{ab}$ and extrinsic curvature $K^{ab}$
of a spacelike hypersurface $\Sigma\hookrightarrow M$ to ``internal
spacetime coordinates'' $X^{A}$, their canonical conjugates $P_{A}$,
and the coordinates on the partially reduced phase space $(\phi^{r},\ p_{r})$.
Using the term ``gauge transformation'' to denote a transformation generated
by the momentum constraints, one might hope for the following: (1) for
arbitrary $(X^{A},\phi^{r},p_{r})$ there exists unique $P_{A}$ such that
the constraints are satisfied; (2) the embedding $\Sigma\hookrightarrow M$
is determined by the functions $X^{A}$, which are otherwise pure gauge; and
(3) the functions $(\phi^{r},p_{r})$ may be regarded as coordinates on a gauge
invariant space, the partially reduced phase space.  If this may be
achieved then
given a spacetime (a manifold with Lorentzian metric satisfying
Einstein's equations), and a canonical transformation \rref{z1},
any foliation into spacelike leaves may be determined
by fixing a one parameter family of functions $X^{A}(t)$.  Such a one
parameter family $X^{A}(t)$ will be referred to as a time-slicing. For each
fixed $t$, $X^{A}(t)|_{\hbox{\scriptsize\it t fixed}}$ will be referred
to as a slice.

The space of time-slicings is homeomorphic to the space of foliations with
spacelike leaves.  However the two should not be identified; if the
spacetime metric is changed the foliation determined by a time-slicing
$X^{A}(t)$ will change.  For example in the Moncrief-Hosoya-Nakao reduction,
the time coordinate is the (constant) mean curvature.  Given a spacetime,
there is a unique embedding $\Sigma\hookrightarrow M$ such that the mean
curvature $K$ has a particular value, for example $K=3$.  However if the
metric is changed, the embedding for which $K=3$ will change.  So in
gravity a particular time is a slice, which is closely related
to, but not quite the same as, an embedding $\Sigma\hookrightarrow M$.

Classically, evolution from slice to slice does not depend on the
intermediate time-slicing.  Therefore, given a solution, a point in
the partially reduced phase space is uniquely specified for each slice.
This phenomenon is
usually called ``many-fingered time''.  In the language of Section~\ref{Sys},
gravity is a system with global time $S$, where $S$ is the space of
slices.  As noted in this and the previous
section, a {\em Hamiltonian} system does not
emerge unless a time-slicing (preferred time) is specified.  For this
reason, given a slice, a Hamiltonian is {\em not} identified.  We
must be given the slice and the local time-slicing to identify the
Hamiltonian.  This is why it is not clear if Hamiltonians should be
considered as observables in gravity.

It must be noted that if the canonical transformation \rref{z1} is not unique,
then viewing gravity as a system with global time $S$ is unnatural.  A
decomposition into time and physical degrees of freedom must still be
{\em chosen}.  The quantum theories based on different choices may be
inequivalent.  This is known as the multiple choice problem \cite{Kuk,Ish1},
a serious problem which is not addressed in this article.  In this article
it is assumed that a particular canonical transformation \rref{z1} has been
chosen, so that the problem known as the functional evolution
problem \cite{Kuk,Ish1} is addressed.

As described, the above conclusions depend on a positive resolution of the
global time problem, that is the existence of the canonical transformation
\rref{z1}.  However these are complex issues, and the above is meant to be
only a cursory introduction, a motivation for viewing gravity as a system
without preferred time parameter as defined in Section~\ref{Sys}.  So let
us finish this section by stating clearly what we will assume in the
remainder of this work.
\begin{asu}
\label{Gti}
Gravity is a system with global time $S$ as defined in
Definition~\ref{dgen} (perhaps not canonically).  For a manifold of the form
$M=\Sigma\times \IR^{+}$,
$S$ is homeomorphic to the space of spacelike embeddings
$\Sigma\hookrightarrow M$ for a particular spacetime $(M,g)$.
\end{asu}
The remainder of this paper will refer to the partially reduced phase space
simply
as the phase space, with the understanding that only systems with global
time $\Gamma$ (as defined in Definition~\ref{dgen}) are considered.

\section{Progress Report}
\label{Pro}

The achievement so far has been to replace the question,
\begin{quote}
Can we achieve consistent evolution in the Heisenberg picture for a
quantum theory without preferred time parameter (for example
gravity)?
\end{quote}
with the questions:
\begin{enumerate}
\item Is quantization after evolution equivalent to Heisenberg equation
evolution when a preferred time has been chosen?
\item Can we carry out quantization after evolution for the theory in question?
\end{enumerate}

If question number 2 is answered affirmatively, but question number 1
unanswered, then we may achieve consistent evolution of observables
defined  at specific global times with the correct classical limit.
However, the evolution may not be generated by a Hamiltonian which
has the correct classical limit.  Here it becomes important to know
if the Hamiltonian should be considered an observable.

\part{Quantization After Evolution}
\label{p2}

In part II, an assumption
concerning the existence of a ``classical limit'' of an algebra of $\hbar$
dependent operators on Hilbert space will be spelled out and used to define
quantization after evolution, as well as ordinary Heisenberg picture quantum
mechanics.  The question of equivalence between the two will be discussed.
Finally, quantization after evolution will be recast in the light of quantizing
canonical transformations, with a view toward the question of existence to be
discussed in part III.

\section{The Classical Limit}
\label{Cla}

It is generally assumed that quantum mechanics reduces to classical
mechanics in some appropriate limit, usually termed ``$\hbar$ goes to zero''.
This is called the correspondence principle.  Operators which arise from
quantization should depend on $\hbar$ in an appropriate way, and there should
be a definition of a limiting process, call it $\lim_{\hbar\rightarrow0}$,
which tests this appropriateness.  The following assumption is motivated
by the work of Werner and Berezin \cite{Wer,Ber}.
\begin{asu}
\label{Lim}
There exists a classical limit, $\lim_{\hbar\rightarrow0}$,
which satisfies the following:
\begin{enumerate}
\item There is an algebra $\hat{A}$ of $\hbar$-sequences,
that is $\hbar$-dependent operators on an appropriate Hilbert space,
which converges to an abelian algebra
whose elements are identified with functions on phase space:
\begin{description}
\item[Convergent Set] For $\hat{a}\in\hat{A},\ \lim_{\hbar\rightarrow0}
(\hat{a})=$
complex valued function on phase space.
\item[Product Law] For $\hat{a}_{1},\ \hat{a}_{2}\in\hat{A},\ \
\lim_{\hbar\rightarrow0}(\hat{a}_{1}\hat{a}_{2})=(\lim_{\hbar\rightarrow0}
(\hat{a}_{1}))(\lim_{\hbar\rightarrow0}(\hat{a}_{2})).$
\end{description}
\item The subset $\hat{A}_{\hbox{\scriptsize\it diff}}\subset\hat{A}$
of ``differentiable-$\hbar$-sequences'', that is $\hbar$-sequences
which converge to
differentiable functions, satisfies the following:
\begin{description}
\item[Lie Bracket Convergence] For $\hat{a}_{1},\ \hat{a}_{2}\in\hat{A}
_{\hbox{\scriptsize\it diff}}
,\ \  \lim_{\hbar\rightarrow0}(\frac{1}{i\hbar}[\hat{a}_{1},\ \hat{a}_{2}])=
\{\lim_{\hbar\rightarrow0}\hat{a}_{1},\ \lim_{\hbar\rightarrow0}\hat{a}_{2}\}
_{pb}$.
\end{description}
\item $\lim_{\hbar\rightarrow0}$ induces a locally-sectionable linear
mapping of the space of Hermitian differentiable-$\hbar$-sequences, to the
space of differentiable real valued functions on phase
space:
$$\rho:\hat{R}_{\hbox{\scriptsize\it diff}}\rightarrow
R_{\hbox{\scriptsize\it diff}},\ \ \rho=
\limh|_{\hat{R}_{\hbox{\scriptsize\it diff}}}.$$
\end{enumerate}
\end{asu}

Locally-sectionable means that there is an open covering $\{U_{i}\}$ of
$R_{\hbox{\scriptsize\it diff}}$, such that for each $U_{i}$ there is a
continuous
mapping $s_{i}:U_{i}\rightarrow\hat{R}_{\hbox{\scriptsize\it diff}}$,
with $\rho\circ s_{i}=I$.
That is, it means that locally it is possible to continuously lift real valued
differentiable functions on phase space to Hermitian operators with the
correct classical limit.

If an operator is not in $\hat{A}$ we shall say that it does not converge, or
is not a convergent $\hbar$-sequence, etc.  Most notably, the unitary
evolution operators of quantum mechanics cannot converge.  However, in
Section~\ref{Uni},
we shall define a sense in which unitary transformations converge to
canonical transformations.

At present, there is no generally accepted implementation of $\limh$.
However, in the finite dimensional case, there are mappings known as the
Wigner-Weyl quantization and dequantization maps \cite{Wer}.
Taking the Weyl symbol \cite{Ber} of an operator
provides a one-to-one mapping of a certain class of operators
$\hat{A^{\prime}}$ to functions on phase space.  For example, if the phase
space is two dimensional the mapping is
\beq
\hat{\obs}\mapsto \frac{1}{2}\frac{\langle q | \hat{\obs} |p\rangle}
{\langle q | p\rangle}+\frac{1}{2}\frac{\langle p | \hat{\obs} |q\rangle}
{\langle p | q\rangle},
\label{z2}
\eeq
and we find
$$\hat{q}\mapsto q,\ \hat{p}\mapsto p,\ \hat{q}\hat{p}\mapsto qp+
\frac{i\hbar}{2},\
\hat{p}\hat{q}\mapsto qp-\frac{i\hbar}{2}\ \ etc.$$
By taking the ordinary limit as $\hbar$ goes to zero of the Weyl
symbols, we get a mapping which satisfies the product law and Lie bracket
convergence for operators in $\hat{A}^{\prime}$.  This operation is generally
regarded as giving the right classical limit for operators in
$\hat{A}^{\prime}$. Because the Weyl symbol-operator correspondence is
one-to-one and continuous, with real valued functions corresponding to
Hermitian
operators, we see  that item 3 of Assumption~\ref{Lim} is satisfied at least
for the subset $\hat{A}^{\prime}\cap\hat{R}_{\hbox{\scriptsize\it diff}}
\subset\hat{R}_{\hbox{\scriptsize\it diff}}$ and
its Weyl symbols.  Hence, if all the functions we cared about were Weyl symbols
of operators in $\hat{A}^{\prime}$, we could proclaim Assumption~\ref{Lim}
satisfied
and $\hat{A}=\hat{A}^{\prime}$.

Actually the preceding paragraph is not rigorous.  What has been achieved,
as I understand it (see \cite{Ber}), is to set up a one to one
correspondence between the functions
$f\in C^{\infty}(\IR^{2n})$ for which there
exist fixed real numbers $C$ and $m$ such that
$$|\partial_{q}^{\alpha}\partial_{p}^{\beta}f(q,p)|\leq C
(1+|q|+|p|)^{m}$$
for all $\alpha$ and $\beta$, and a class of operators that have a dense
domain in $L^{2}(\IR^{n})$.  Equation \rref{z2} is at least valid for
polynomial functions.

Recently Werner \cite{Wer} has proposed a definition for $\limh$ in the
finite dimensional
case which purports to agree with the accepted folklore and be valid on the
set of bounded operators.  Werner shows that his construction satisfies item 1
and item 2 of Assumption~\ref{Lim}.  Item 3 is not explored.  It is quite
possible
that in the case of (2+1)-dimensional gravity, where the reduced phase space
is finite dimensional, Werner's work will provide the hoped for definition of
the classical limit, and allow all the assumptions
(except Assumption~\ref{Gti}) of this paper to be tested.

The following lemma brings out the importance of item 3  of
Assumption~\ref{Lim}, and
is required in part III.  It is also tacitly used in the definitions that
follow, although they could be modified to do without it.
\begin{lem}
\label{Lem}
If the topology on the space of differentiable real valued functions on
phase space is paracompact, and if Assumption~\ref{Lim}  holds, then the
differentiable real valued functions may be continuously lifted to
hermitian operators on Hilbert space with the correct classical limit.
\end{lem}

{\bf Proof}:  Let $R_{\hbox{\scriptsize\it diff}}$
be the space of real valued differentiable
functions with a paracompact topology.
Let $\hat{R}_{\hbox{\scriptsize\it diff}}$ be the space
of hermitian differentiable-$\hbar$-sequences.  We need to
show that the bundle over $R_{\hbox{\scriptsize\it diff}}$,
$$\rho:\hat{R}_{\hbox{\scriptsize\it diff}}\rightarrow
R_{\hbox{\scriptsize\it diff}},
\ \ \rho=\limh|_{\hat{R}_{\hbox{\scriptsize\it diff}}},$$
has a section.

Let $\hat{r}$ and $\hat{r}'$ be in $\rho^{-1}(r)$ where $r$ is in
$R_{\hbox{\scriptsize\it diff}}$,
then
$$\rho(\lambda\hat{r}+(1-\lambda)\hat{r}')=r.$$
Therefore $\rho^{-1}(r)$ is a convex domain.

If $R_{\hbox{\scriptsize\it diff}}$ is paracompact, then a
locally-sectionable bundle over $R_{\hbox{\scriptsize\it diff}}$
with convex fibers has a section.  A paracompact space
has a partition of unity subordinate to any locally finite covering \cite{Cho}.
Therefore choose a sufficiently fine covering $\{U_{i}\}$ of
$R_{\hbox{\scriptsize\it diff}}$,
and a partition of unity $\{f_{i}\}$ subordinate to it.  Because the bundle
over $R_{\hbox{\scriptsize\it diff}}$ is locally sectionable we can construct
mappings
$$\sigma_{i}:U_{i}\rightarrow \hat{R}_{\hbox{\scriptsize\it diff}},
\ \ \rho\circ\sigma_{i}=I|_{U_{i}}
.$$
Then, because the fibers are convex,
$$\sum_{i}f_{i}\sigma_{i}:R_{\hbox{\scriptsize\it diff}}
\rightarrow \hat{R}_{\hbox{\scriptsize\it diff}}$$
is a section.\\
{\bf QED}

 From a physics point of view, the paracompactness requirement is innocuous.
Any topology which may be defined by a metric (metrizeable topology) is
paracompact \cite{Cho}.  If for some reason one wanted to drop the
paracompactness requirement, the word ``locally'' would have to be removed
 from item 3 of Assumption~\ref{Lim} in what follows.

\section{Heisenberg Picture Quantum Mechanics}
\label{Hei}

The general concept of $\limh$ will now be used to make various
definitions. The first one is Heisenberg picture quantum mechanics.
\begin{Def}
\label{Dhe}
Let $\obs_{H(t)}$ be the phase space observable corresponding to the
Hamiltonian at time $t$.
Heisenberg picture quantum mechanics is a mapping,
\begin{eqnarray*}
HQ:\Obs\sqcup \{(\obs_{H(t)},t)\ |\ t\in\IR\}&\rightarrow&
\{\hbox{\it Hermitian\ operators}\}\\
\obs\mapsto\hat{\obs}(t_{0})&,&\ (\obs_{H(t)},t)\mapsto\hat{H}(t),
\end{eqnarray*}
such that:
\begin{enumerate}
\item $\limh(\hat{\obs}(t_{0}))=\obs,\ \ \limh(\hat{H}(t))=\obs_{H(t)},$
\item For a certain ``preferred'' Lie subalgebra of phase space
observables $\{\alpha\}
\subset\Obs$, $HQ$
is a lie algebra isomorphism:
$$HQ(\{\beta,\ \alpha\}_{pb})=\frac{1}{i\hbar}[\hat{\beta}(t_{0}),\
\hat{\alpha}(t_{0})],\ \ \beta,\ \alpha\in\{\alpha\},$$
\end{enumerate}
together with the Heisenberg evolution equation,
$$\frac{d\hat{\obs}}{dt}(t)=\frac{1}{i\hbar}[\hat{\obs}(t),\ \hat{H}(t)],$$
which determines $\hat{\obs}(t)$ for $t>t_{0}$.
\end{Def}

Note the use of the disjoint union to adjoin the set of Hamiltonians to the
set of phase space observables.  A time-dependent Hamiltonian is not properly
a function on phase space.  It only gives rise to a phase space observable
$\obs_{H(t)}$ (function on phase space) when evaluated at a
particular time.  As mentioned earlier (Section~\ref{Gra}), in
geometrodynamics the situation is
even worse.  The Hamiltonian is not defined at a time (that is on a slice),
but only for a local time-slicing.  However, it is still true that given a
time and a local time-slicing, a function on phase space is given, the phase
space observable corresponding to the Hamiltonian for the time and local
time-slicing.  By considering the set of Hamiltonians as separate from the
set of phase space observables,
we are allowing the Hamiltonian to lift to an operator which
differs from the operator for the
related phase space observable by a term with zero classical limit.  Also,
because we are considering the pairs $(\obs_{H(t)},t)$,
Hamiltonians at different times, but which correspond to the
same phase space observable may lift to operators that differ by a term with
zero classical limit.  Hence
some of the structure of the classical theory, specifically the relationship
between observables and generators of evolution, is lost in the quantum theory.
In run-of-the-mill quantum mechanics, the set of Hamiltonians is regarded as
a subset of the space of phase space observables.

In determining whether or not Definition~\ref{Dhe} is reasonable, one should
consider
that we always give up structure when quantizing.  For example, Van Hove's
theorem implies that we can never completely preserve the Lie algebra given
by the Poisson bracket.  Only a very small subalgebra can be preserved
\cite{geo}.
Actually, when quantizing a theory, we generally consider only a few
phase space observables.
So to evaluate the
relation of Definition~\ref{Dhe} to experimentally tested aspects of quantum
mechanics, let us restrict our attention to a discrete
subset of phase space observables, $\{q,p\}\subset\Obs$, which parametrize
phase space.  $HQ$ is replaced by the restricted mapping $HQ^{\prime}$:
\begin{eqnarray*}
HQ^{\prime}:\{q,p\}\sqcup\{(\obs_{H(t)},t)\ |\ &t&\in\IR\}\rightarrow
\{Hermitian\ operators\}\\
q\mapsto\hat{q}(t_{0}),&\ &p\mapsto\hat{p}(t_{0}),\ (\obs_{H(t)},t)
\mapsto\hat{H}(t).
\end{eqnarray*}
It is rare that the function on phase space corresponding to
the Hamiltonian is
$q$ or $p$.  So the concern that a phase space observable and a
Hamiltonian
which corresponds to the same function on phase space may map to different
operators is almost moot.  The only remaining distinction between
Definition~\ref{Dhe}
and run-of-the-mill quantum mechanics is that even if the classical
Hamiltonian is not time-dependent,  Definition~\ref{Dhe} allows the operator
Hamiltonian to be time-dependent.  So the phrase to remember is that
Definition~\ref{Dhe}
allows time dependent operator orderings for Hamiltonians.  In a theory
where a time-independent classical Hamiltonian is the exception, not the rule,
time-dependent operator ordering does not seem a big liberty.

\section{Quantization After Evolution}
\label{Qua}

The mapping $\limh$ can be used to give an equivalent formulation
of the classical
limit in terms of the solution space observables.  We choose an isomorphism
between the phase space $PS$ and the solution space $SS$ by choosing an
``initial'' global time, $\tau_{0}$, and mapping a point $\eta$ in
phase space to the {\em solution} which has the ``initial data'' $\eta$ at
time
$\tau_{0}$ (note that Assumption~\ref{Gti} guarantees that this mapping is
an isomorphism):
\beq
I_{\tau_{0}}: PS \rightarrow SS.
\label{g0}
\eeq
$I_{\tau_{0}}$ induces the obvious isomorphism $I_{\tau_{0*}}$
between functions on phase space and functions on solution space:
\begin{eqnarray}
I_{\tau_{0*}}:\{functions\ on\ PS\}&\rightarrow&
\{functions\ on\ SS\}\nonumber\\
\obs\mapsto\obs_{\tau_{0}}&,&\ \ \obs(\eta)=\obs_{\tau_{0}}(I_{\tau_{0}}\eta).
\label{g1}
\end{eqnarray}
Of course $I_{\tau_{0}*}$ is just the mapping of Definition~\ref{evo}
(Section~\ref{Sys})  for
$\tau=\tau_{0}$;  that is $I_{\tau_{0}}\eta (\tau_{0})=\eta$.
Finally define the limit $\limht$ to be the composition of
$\limh$ with $I_{\tau_{0*}}$:
\begin{eqnarray}
\limht:\{convergent\ operators\}&\rightarrow&\{
functions\ on\ SS\}\nonumber\\
\limht&=&I_{\tau_{0}*}\comp\limh.
\label{g3}
\end{eqnarray}

Now we are in a position to define quantization after evolution:
\begin{Def}
\label{Dqu}
A  quantization after evolution is a  mapping,
$$QAE:AES\rightarrow\{Hermitian\ operators\},\ \ (\obs_{\tau},\tau)
\mapsto\hat{\obs}
(\tau),$$
differentiable in $\tau$, such that:
\begin{enumerate}
\item It has the correct classical limit, that is
$$\limht(\hat{\obs}(\tau))=\obs_{\tau}.$$
\item For a certain ``preferred'' Lie subalgebra of solution space
observables at a fixed time $\tau_{0}$,
$$O\subset\{(\obs_{\tau},\tau)\ |\ \tau=\tau_{0}\}
\subset AES,$$
$QAE$ is a lie algebra isomorphism:
$$QAE((\{\alpha_{\tau_{0}},\ \beta_{\tau_{0}}\}_{pb},\tau_{0}))=
\frac{1}{i\hbar}[\hat\alpha(\tau_{0}),
\ \hat\beta(\tau_{0})],\ \ (\alpha_{\tau_{0}},\tau_{0}),\
(\beta_{\tau_{0}},\tau_{0})\in O.$$
\item $QAE$ respects the causal structure $AES\simeq\Obs\times\Gamma$
provided by the bijection $B$ \rref{bij}.  The natural (and known from
quantum mechanics) meaning of this is:
\begin{eqnarray*}
For\ all\ \tau_{1},\tau_{2}\in\Gamma,\ there&\ & exists\ a\ unitary\
operator\  \hat{U}(\tau_{1},\tau_{2})\ such\ that:\\
\hat{\obs}(\tau_{2})&=&\hat{U}(\tau_{1},\tau_{2})\hat{\obs}(\tau_{1})
\hat{U}^{\dag}(\tau_{1},\tau_{2}).
\end{eqnarray*}
\end{enumerate}
\end{Def}
Definition~\ref{Dqu} depends on $\tau_{0}$, because equation \rref{g3}
depends on $\tau_{0}$. However, the dependence is trivial  (and is therefore
not reflected in the notation, $\limht$).  The relation between phase space
and solution space is simply shifted by a canonical transformation of
solution space (or a canonical transformation of phase space).

To understand the relationship between Definition~\ref{Dqu} and the brief
description
given in Section~\ref{Qua1}, consider the following diagram:
$$
\begin{array}{ccc}
\left\{\stackrel{\textstyle{quantized}}{\textstyle{solution\ space}}\right\}&
\stackrel{\textstyle{D}}{\Longrightarrow}&
\left\{\stackrel{\textstyle{quantized}}{\textstyle{AES}}\right\}\\
C\uparrow&\ &B\uparrow\\
\left\{\textstyle{solution\ space}\right
\}&
\stackrel{\textstyle{A}}{\Longrightarrow}&
\left\{\textstyle{AES}\right\}
\end{array}
$$

The single arrow $C$ denotes quantizing the solution space.  This involves
choosing a preferred Lie subalgebra of solution space observables
to be preserved.  The double arrow $D$ denotes quantization after evolution
by substituting the operators obtained from $C$ into the functions of the
$AES$ (see Figures 1 and 2) and attempting to find operator orderings such
that items 2 and 3 of Definition~\ref{Dqu} are satisfied.  Due to results
such as Van Hove's theorem this may not be possible.  For example, preservation
of the preferred Lie subalgebra  of the process $C$ may be incompatible with
preservation of the preferred Lie subalgebra of the quantization after
evolution.  Hence the path $D\circ C$, which is the path described in
Section~\ref{Qua1}, is not passable for all quantizations of solution space
$C$.  In reference \cite{Car1}, Carlip had to follow this path, using
Witten's parametrization of solution space, to show that Witten's quantization
of (2+1)-gravity was equivalent to the Moncrief-Hosoya-Nakao quantization.

The double arrow $A$ denotes construction of the already evolved classical
system.  This involves defining functions on solution space.
The single arrow $B$ denotes quantization after evolution any way possible.
For example if the space of global time is $\IR$, we can always use
evolution under the Heisenberg equations of motion (except see
Assumption~\ref{Ahc}).
The path $B\circ A$ is the path described in this section.  If the space of
global time is $\IR$, it is passable whenever the corresponding
Heisenberg picture
quantization (Definition~\ref{Dhe}) is possible.
It was briefly described by Kucha\v{r} in reference \cite{Kuk}.

\section{Comparison of Quantization After Evolution and Heisenberg
Picture Quantum Mechanics}
\label{Com}

In the case $\Gamma=\IR$, or when a preferred time has been chosen, we can
compare quantization after evolution with Heisenberg picture quantum
mechanics.  This is done by using the bijection $B:\Obs\times\Gamma
\rightarrow AES$ \rref{bij},
to map Heisenberg picture quantum mechanics into a quantization of the
already evolved classical system ($AES$).  The result is the following
equivalent definition of Heisenberg picture quantum mechanics:
\begin{Def}[equivalent to Definition~\ref{Dhe}]
\label{Deq}
Let $(H_{t},t)$ be the solution space observable corresponding to the
Hamiltonian
at time t, that is $(H_{t},t)=B(\obs_{H(t)}, t)$.
Heisenberg picture quantum mechanics is a mapping
\begin{eqnarray*}
HQA:AES\sqcup\{(H_{t},t)\ |\ t\in\IR\}
&\rightarrow&\{Hermitian\ operators\},\ defined\ by\\
(\obs_{t},t)\mapsto\hat{\obs}(t),\ (H_{t},t)&\mapsto&\hat{H}(t),
\end{eqnarray*}
such that:
\begin{enumerate}
\item $\limht(\hat{H}(t))=H_{t}$, and for a fixed time $t_{0}$
$\limht(\hat{\obs}(t_{0}))=\obs_{t_{0}}$.
\item $\frac{d\hat{\obs}}{dt}(t)=\frac{1}{i\hbar}[\hat{\obs}(t),\
\hat{H}(t)]$.
\item For a certain ``preferred'' Lie subalgebra of the solution space
observables at time $t_{0}$,
$$O\subset\{(\obs_{t},t)\ |\ t=t_{0}\}\subset AES,$$
$HQA$ is a lie algebra isomorphism:
$$HQA((\{\alpha_{t_{0}},\ \beta_{t_{0}}\}_{pb},t_{0}))=
\frac{1}{i\hbar}[\hat{\alpha}(t_{0}),\
\hat{\beta}(t_{0})],\ \ (\alpha_{t_{0}},t_{0}),\ (\beta_{t_{0}},t_{0})\in O.$$
\end{enumerate}
\end{Def}

To see that Definition~\ref{Deq} is equivalent to Definition~\ref{Dhe}, we
must check that the operators of Definition~\ref{Deq} have the right
classical limit under $\limh$.  Specifically we require
\begin{eqnarray*}
\limh\hat{\obs}(t_{0})&=&\obs\ \ and,\\
\limh\hat{H}(t)&=&\obs_{H(t)}.
\end{eqnarray*}
Applying the identity in the form $I^{-1}_{t_{0}*}I_{t_{0}*}$ we find
\begin{eqnarray}
I^{-1}_{t_{0}*}I_{t_{0}*}\limh \hat{\obs}(t_{0})&=&I^{-1}_{
t_{0}*}\limht\hat{\obs}(t_{0})=I^{-1}_{t_{0}*}\obs_{t_{0}}=\obs\ \ and,
\label{star}\\
I^{-1}_{t_{0}*}I_{t_{0}*}\limh \hat{H}(t)&=&I^{-1}_{
t_{0}*}\limht\hat{H}(t)=I^{-1}_{t_{0}*}H_{t}.\label{tot}
\end{eqnarray}
The last equality of Equation \rref{star} follows because $I_{t_{0}*}$
is the mapping of Definition~\ref{evo} for $t=t_{0}$ (see section~\ref{Qua}).

It remains to show that $I^{-1}_{t_{0}*}H_{t}=\obs_{H(t)}$.  Again from the
definition of $I_{t_{0}*}$ and Definition~\ref{evo} (Section~\ref{Sys}) we have
\begin{eqnarray}
I_{t_{0}*}\obs_{H(t)}(\xi)&=&\obs_{H(t)}(\xi(t_{0}))\ \ and,\nonumber\\
H_{t}(\xi)&=&\obs_{H(t)}(\xi(t)).
\label{tt}
\end{eqnarray}
Evolution of $\obs_{H(t)}(\xi(t))$ is given by the Hamiltonian equations
of motion:
\begin{eqnarray*}
\frac{d\obs_{H(t)}}{dt}-\frac{\partial\obs_{H(t)}}{\partial t}&=&
\{\obs_{H(t)},\ \obs_{H(t)}\}_{pb}=0,\\
so\ \ \frac{d\obs_{H(t)}}{dt}&=&\frac{\partial\obs_{H(t)}}{\partial t}.
\end{eqnarray*}
Therefore $\obs_{H(t)}(\xi(t_{0}))=\obs_{H(t)}(\xi(t))$ and from \rref{tt}
we have
$$H_{t}=I_{t_{0}*}\obs_{H(t)},$$
which completes the demonstration that Definition~\ref{Deq} and
Definition~\ref{Dhe} are equivalent.

Comparing Definition~\ref{Dqu} with Definition~\ref{Deq} we see that
Heisenberg picture quantum mechanics is equivalent to quantization
after evolution (at least if we restrict our attention to differentiable
functions) if the following assumptions hold:
\begin{asu}
\label{Ahc}
Evolution under the Heisenberg equations of motion implies that the
correct classical
limit is obtained for $t>t_{0}$.  That is
$\limht(\hat{\obs}(t))=\obs_{t}$ for all $t>t_{0}$.
\end{asu}
\begin{asu}
\label{Ach}
Differentiable unitary evolution with the correct classical
limit implies that
$$\frac{d\hat{\obs}}{dt}(t)=\frac{1}{i\hbar}[\hat{\obs}(t),\ \hat{H}(t)]$$
for some $\hat{H}(t)$ with the correct classical limit.
\end{asu}
If Assumption~\ref{Ahc} holds, then Heisenberg picture quantum
mechanics is contained in quantization after evolution.  If
Assumption~\ref{Ach}
holds, then quantization after evolution is contained
in Heisenberg picture quantum mechanics.

Assumptions \ref{Ahc} and \ref{Ach} can in principle be checked once we have
a physically accepted definition for $\limh$.  Using his definition,
Werner has shown that Assumption~\ref{Ahc} holds for time-independent
Hamiltonians.  It may be somewhat over cautious to refer to
Assumption~\ref{Ahc}
as an assumption.  It is, after all, a general tenet of quantum mechanics.
Never the less we will continue to do so.

If Assumption~\ref{Lim} (Section~\ref{Cla}) holds, then Assumption~\ref{Ach}
holds if the Hamiltonian is in
$\hat{A}_{\hbox{\scriptsize\it diff}}$ and the
classical limit commutes with the time derivative.  The only thing we can say
here is that if the classical limit commutes with the time derivative
then it seems probable that the Hamiltonian is constrained to be in
$\hat{A}_{\hbox{\scriptsize\it diff}}$.  To understand this statement
consider application of
the classical limit to the operator
evolution equation:
$$\limh\left(\frac{d\hat{\obs}}{dt}\right)(t)=\limh\left(\frac{1}{i\hbar}
[\hat{\obs}(t),\hat{H}(t)]\right),
\ \ \hat{H}(t)=-i\hbar\hat{U}^{-1}\frac{d\hat{U}}{dt}.$$
If we assume that the classical limit commutes with the time derivative we
find
\beq
\{\limh(\hat{\obs}),\ \obs_{H(t)}\}_{pb}=\limh\left(\frac{1}{i\hbar}
[\hat{\obs},\
\hat{H}(t)]\right).
\label{k1}
\eeq
Now write $\hat{H}(t)=\hat{H}^{\prime}(t)+\hat{h}(t)$ where $\limh(\hat{H}
^{\prime}(t))=\obs_{H(t)}$.  If Assumption~\ref{Lim} holds then
$\limh(\frac{1}{i\hbar}
[\hat{\obs}(t),\ \hat{H}^{\prime}(t)])=\{\limh(\hat{\obs}(t)),
\ \obs_{H(t)}\}_{pb}$ and we have
\beq
\limh\left(\frac{1}{i\hbar}[\hat{\obs}(t),\ \hat{h}(t)]\right)=0.
\label{k2}
\eeq
If the Hamiltonian is not in $\hat{A}_{\hbox{\scriptsize\it diff}}$, then
$\limh(\hat{h}(t))$
is not a differentiable function.  In contradistinction to \rref{k2},
it seems likely that if
$\limh(\hat{h}(t))$ is not a differentiable
function, and $\limh (\frac{1}{i\hbar}[\hat{f},\ \hat{g}])=\{\limh (\hat{f}),\
\limh (\hat{g})\}_{pb}$ when $\limh (\hat{f})$ and $\limh (\hat{g})$ are
differentiable (Assumption~\ref{Lim}),
then $\limh(\frac{1}{i\hbar}[\hat{\obs}(t),\ \hat{h}(t)])$ would
be something
like the Poisson bracket with a non-differentiable function, or worse.

\section{Unitary Transformations Over Canonical Transformations}
\label{Uni}

There is another way to look at quantization
after evolution which focuses less on observables and more on unitary
transformations.  In this respect it is better tied to the original
motivation of Section~\ref{Mot}.  To explain this picture, we first define
a bundle of unitary transformations over canonical transformations.
By a ``bundle'' is meant a triple, $(Y,X,\rho)$, consisting of two
two topological spaces $Y$ and $X$ and a continuous surjective
mapping $\rho:Y\rightarrow X$ \cite{Jam}.

Let $U$ be the connected component of the identity of the space of
canonical transformations of solution space.
Let $\hat{U}$ be the space of unitary transformations of Hilbert space
such that for $\hat{u}$ in $\hat{U}$ there exists $u$ in $U$ with
$\limht(\hat{u}(\hat{f}))=u(\limht(\hat{f}))$ for all
differentiable-$\hbar$-sequences $\hat{f}$.

We wish to define topologies and differential structures on $U$ and
$\hat{U}$ consistent with the topologies and differential structures
assumed to exist on the space of differentiable functions
on solution space $A_{\hbox{\scriptsize\it diff}}$, and the space of
differentiable-$\hbar$-sequences on Hilbert space
$\hat{A}_{\hbox{\scriptsize\it diff}}$,
respectively.
Let $\hat{V}$ and $V$ be arbitrary open sets in $\hat{A}
_{\hbox{\scriptsize\it diff}}$ and $A_{\hbox{\scriptsize\it diff}}$
, respectively.  Let $\hat{f}$ and $f$ be arbitrary elements of
$\hat{A}_{\hbox{\scriptsize\it diff}}$
and $A_{\hbox{\scriptsize\it diff}}$, respectively.
Define an open set in $U$ labeled by
$f$ and $V$ to be
\beq
U_{f,V}=\{u\in U|u(f)\in V\}.
\eeq
Define an open set in $\hat{U}$ labeled by $\hat{f}$ and $\hat{V}$ to be
\beq
\hat{U}_{\hat{f},\hat{V}}=\{\hat{u}\in \hat{U}|\hat{u}(\hat{f})\in \hat{V}\}.
\eeq
For all $f,\hat{f},V,\hat{V}$
these open sets generate topologies on $U$ and $\hat{U}$.
These topologies ensure that:
\begin{enumerate}
\item  Given any $\hat{f}$, a continuous
path in $\hat{U}$  maps to a continuous path in $\hat{R}
_{\hbox{\scriptsize\it diff}}$ by its
action on $\hat{f}$.
\item Given any $f$, a continuous
path in $U$  maps to a continuous path in
$R_{\hbox{\scriptsize\it diff}}$ by its action on $f$.
\end{enumerate}
Define the differentiable
structures on $\hat{U}$ and $U$ similarly.  That is $C^{1}$ paths map to
$C^{1}$ paths.

To complete the definition of the bundle, $(\hat{U},U,\rho)$, define the
obvious mapping
\begin{eqnarray}
\rho:\hat{U}&\rightarrow& U\ \ by\nonumber\\
\hat{u}\mapsto u\ \ &if&\ \ \limht(\hat{u}(\hat{f}))=u(\limht(\hat{f})).
\end{eqnarray}
The inverse images $\rho^{-1}(u)$ for $u$ in $U$ are diffeomorphic
to each other, since they are related by unitary
transformations. I shall call them fibers. Points within a
fiber are related by unitary
transformations which lie in the fiber over the identity.  These are the only
unitary transformations enacted by unitary operators which can converge
under $\limh$ because if $\hat{U}_{n}$ is a convergent unitary
operator then by the product law of Assumption~\ref{Lim} (Section~\ref{Cla}),
$$\limh(\hat{U}_{n}\hat{f}\hat{U}_{n}^{\dagger})=
\limh(\hat{f})\limh(\hat{U}_{n}\hat{U}_{n}^{\dagger})=\hat{f}.$$

\section{Quantizing Canonical Transformations}
\label{Qua2}

The alternative picture of quantization after evolution we have in mind may
be described as quantizing canonical transformations  \cite{And}.  If the
bundle, $(\hat{U},U,\rho)$, has a  section, then the symmetries
of the classical theory can be reproduced exactly in the quantum theory.
That is the canonical transformations
can be lifted continuously to unitary transformations with the correct
classical limit.  It is tempting to conclude from  Van Hove's theorem that
that this cannot be done.  However this is not clear, we are not requiring
that unitary transformations be generated by observables (see
Section~\ref{Hei}).
Regardless, in quantization after evolution we are only
interested in the canonical transformations which generate time evolution.

In order to limit consideration to canonical transformations which generate
time evolution, define a mapping of the space of global time $\Gamma$
into the base space $U$ of canonical transformations.  Fix an initial
time $\tau_{0}$, and map it to the identity transformation.  Then associated
to every other time $\tau$ is a canonical transformation defined by
classical evolution from $\tau_{0}$ to $\tau$.\footnote{In gravity,
simply take any time-slicing connecting the ``initial'' slice $\sigma_{0}$
and the slice $\sigma$.  (In gravity the space of global time is the
space of
slices.  The label $\sigma$ is used to distinguish this particular case
 from the general case.)  Since the
classical theory is independent of time-slicing, the resulting
canonical transformation is too.  If $\sigma_{0}$ and $\sigma$ are overlapping
slices, evolve from $\sigma_{0}$ to $\sigma^{\prime}$, which does not
overlap either, and then from $\sigma^{\prime}$ to $\sigma_{0}$.  Actually
this can be viewed as a piecewise differentiable choice of preferred time
which is not a time-slicing.  Clearly there is also a differentiable choice
of preferred time connecting $\sigma_{0}$ and $\sigma$.}
This defines the mapping $T_{\Gamma}$
and gives rise to the following diagram:
\beq
\begin{array}{ccl}
\ &\ &\hat{U}\\
\ &\ &\downarrow \rho\\
\Gamma&\rightarrow&U\ \ .\\
\ &T_{\Gamma}&\ \
\end{array}
\label{m3}
\eeq
Together with Definition~\ref{dgen} (Section~\ref{Sys}),
the topology and differential structure on $U$ ensure
that $T_{\Gamma}$ is differentiable.

The pullback bundle over $\Gamma$, $(P,\Gamma,\lambda)$, is defined as
\begin{eqnarray}
&P&=\{(\tau,\hat{u})\in\Gamma\times\hat{U}\ |\ T_{\Gamma}(\tau)=\rho(\hat{u})\}
\nonumber\\
&\lambda&:P\rightarrow\Gamma,\nonumber\\
&(&\tau,\hat{u})\mapsto \tau.
\label{m}
\end{eqnarray}
A section in the pullback bundle over $\Gamma$ is a continuous mapping
of the space of global time to unitary transformations which has
the correct classical limit in the sense of Section~\ref{Uni}.  That is,
it is a commutative diagram
\beq
\begin{array}{ccl}
\ &\ &\hat{U}\\
\ &\nearrow&\downarrow \rho\\
\Gamma&\rightarrow&U\ \ .\\
\ &T_{\Gamma}&\ \
\end{array}
\label{m2}
\eeq
Now assume that a mapping of the solution space observables corresponding
to time $\tau_{0}$, to operators, which satisfies conditions 1 and 2 in
the definition of quantization after evolution (Definition~\ref{Dqu},
Section~\ref{Qua}) has been
chosen. By Lemma~\ref{Lem} (Section~\ref{Cla}) this may always be done.
The remaining
freedom in choice of a quantization after evolution is the choice of a $C^{1}$
section in the pullback bundle over $\Gamma$.  The existence of a
$C^{1}$ section
is equivalent to the existence of a quantization after evolution.
\addtocounter{footnote}{-2}

\part{Consistent Evolution With Different Time-Slicings}
\label{p3}

It will serve as a good introduction to sketch the most obvious approach to
showing that consistent evolution may be achieved, although in the end we will
not be able follow through with this approach, and will settle for a slightly
weaker result.
Consider a spacetime manifold of the form $M=\Sigma\times\IR^{+}$, and
take $\sigma_{0}=\Sigma\times 0$.
Assume we can show that each slice $\sigma$ can be connected to
$\sigma_{0}$ by a particular time-slicing, such that the variation
of the time-slicings with $\sigma$ is continuous.\footnote{The choice of such
continuous family of time-slicings may be quite arbitrary.}
By Assumption~\ref{Lim}  and Lemma~\ref{Lem} (both in Section~\ref{Cla}) the
classical Hamiltonians for these time-slicings
can be continuously lifted to operator Hamiltonians
with the right classical limit.
Hence each slice $\sigma$ can be assigned a unitary transformation
by evolving from $\sigma_{0}$ (which is assigned the identity transformation)
to $\sigma$ using the Heisenberg equations of motion.  Because the
time-slicings and Hamiltonians vary continuously with $\sigma$,
the variation of the unitary transformations
with $\sigma$ will be continuous.  Because the unitary transformations are
derived from the Heisenberg equations of motion, they will have the right
classical limit in the sense of Section~\ref{Uni}.
Hence we will have shown that a section of the pullback bundle over $\Gamma$
exists when $\Gamma$ is the space of slices
(Diagram \rref{m2} with $\Gamma=S=space\ of\ slices$).
If we further assume that the existence of a $C^{0}$ section implies the
existence of a $C^{1}$ section, then we will have shown that a quantization
after evolution of gravity  exists.\footnote{The last assumption can be
described
as an assumption of reasonable compatibility between the topology and
differential structure of the space of operators on Hilbert space.}

The criterion that each slice $\sigma$ can be connected to $\sigma_{0}$
by a particular time-slicing, such that the variation of the time-slicings with
$\sigma$ is continuous, will be called the time-slicing criterion.
It is difficult to prove that the time-slicing criterion holds in general.
Hence, I will not be able to show existence of consistent evolution in
complete generality.
As a substitute I will first offer a non-rigorous physical argument for the
time-slicing criterion in (2+1)-dimensions.  Then, feeling that this is not
enough, I will derive a sufficient condition for the pullback bundle over
$\Gamma$ \rref{m} (when $\Gamma$ is
an arbitrary space of global time)  to have a section \rref{m2}, and
show that the condition is satisfied when $\Gamma$ is  any subspace
of the space of slices parametrized by a (finite or infinite) CW-complex.
This result will essentially mean that if Assumption~\ref{Lim} holds,
then we can have consistent evolution for at least an infinite parameter
family of time-slicings.  Finally I will show that any particular
time-slicing, with evolution defined by any operator ordering of the
associated Hamiltonian, may be included in such an infinite parameter family.

\section{Physical Argument}
\label{Phy}

Although we cannot prove that the time-slicing criterion holds---that is
that each slice $\sigma$ can be connected to
$\sigma_{0}$
by a particular time-slicing, such that the variation of the time-slicings
with $\sigma$ is continuous---we can see the likelihood of this result
 from a simple physical model.  Consider (2+1)-dimensional gravity with
$\Sigma=\Sigma_{g}$, a genus $g$ surface.
The question is mostly one of
topology, not geometry.  So instead of the $M^{2+1}=\Sigma_{g}\times \IR^{+}$
spacetime, consider a thickened genus g surface made of foam rubber.
Further imagine that the piece of foam rubber has been
constructed by gluing together thin genus g shells of foam rubber
(a genus g onion).  This foam rubber object is a ``physical representation''
of the topology of the region of $M^{2+1}$ between $\sigma$ and $\sigma_{0}$
with preferred time-slicing.

Note that we have exchanged the topology on the space of slices
$S$ (which is related to the topology on $U$ via $T_{\Gamma}|_{\Gamma=S}$
\rref{m3}) for the topology placed on a layered piece of foam rubber by
our intuition.
Clearly this is not rigorous.  However, the topology on $S$ is one in
which the spatial metrics induced on the slices (points of $S$) by the
spacetime metric
change continuously.  Locally
the spacetime manifold has the topology of $\IR^{3}$ so that the spacetime
metric is locally a continuous function on $\IR^{3}$.  Since $\IR^{3}$ is
surely the topology of our intuition, our intuition should be applicable
to judging what is meant by a continuous change in slice.  At least it seems
worthwhile to continue.

To see that the time-slicing changes continuously
with $\sigma$, simply poke and prod at the outside of the foam rubber (that is
deform $\sigma$) and imagine the deformation of the interior layered foam
rubber structure (that is deformation of the time-slicing).  Clearly for
each poke and prod we get a unique distribution of the interior foam rubber,
and when we remove our fingers, it springs back to its original distribution.
This suggests that the time-slicing condition is satisfied.  Unfortunately,
we have little experience with $4$-dimensional foam rubber.

\section{Generalized Condition}
\label{Gen}

We will now derive a sufficient condition for the pullback bundle over an
arbitrary space of global time $\Gamma$ \rref{m} to have a section
\rref{m2}.

\begin{cond}
\label{Eco}
Let E be the space of pairs, of points $\tau$ in $\Gamma$, and paths $p$
in $U$ connecting $T_{\Gamma}(\tau)$ to $T_{\Gamma}(\tau_{0})=I$
(see \rref{m3}):
$$E=\{(\tau,p)\ |\ \tau\in\Gamma,\ p\ a\ C^{1}\ path\ in\ U\ connecting\
T_{\Gamma}(\tau)\ to\ I\}.$$
We will say Condition~\ref{Eco} is satisfied if
the bundle, $(E,\Gamma,\sigma)$, defined by
\begin{eqnarray*}
\sigma:E&\rightarrow&\Gamma,\\
(\tau,p)&\mapsto& \tau,
\end{eqnarray*}
has a section.
\end{cond}
If the time-slicing criterion of Section~\ref{Phy} is satisfied, then
Condition~\ref{Eco} is satisfied when $\Gamma$ is the space of slices $S$,
because a time-slicing is a path in $S$ which is mapped to a path in $U$ by
$T_{\Gamma}$.

\begin{theo}
\label{Th1}
If the space of functions on solution space is paracompact, and
if Condition~\ref{Eco}, Assumption~\ref{Lim} (Section~\ref{Cla}), and
Assumption~\ref{Ahc} (Section~\ref{Com}) hold,
then the pullback bundle over $\Gamma$ \rref{m} has a section \rref{m2}.
\end{theo}

The argument for Claim~\ref{Th1} is analogous to the argument for the
time-slicing criterion sketched in the introduction to Part~\ref{p3}.
In view of Lemma~\ref{Lem} (Section~\ref{Cla}), fix a particular continuous
lifting of the
differentiable
real valued functions on solution space to Hermitian operators:
\beq
L:R_{\hbox{\scriptsize\it diff}}\rightarrow \hat{R}
_{\hbox{\scriptsize\it diff}},\ \ \limht\comp L=I.
\eeq
Then consider a $C^{1}$ path in $U$ starting at $T_{\Gamma}(\tau_{0})=I$ and
ending at
the image of another point in $\Gamma$, $T_{\Gamma}(\tau)$:
\beq
p:[0,1]\rightarrow U,\ \ p(0)=I,\ p(1)=T_{\Gamma}(\tau),\ \
\tau\in\Gamma.
\eeq
Together, the mappings $L$ and $p$ define a lifting of $\tau$ in the pullback
bundle over $\Gamma$ as follows.  To each point $t\in[0,1]$ on the path in $U$
is associated a Hamiltonian $H_{pt}$.  Use the lifting $L$ to map it to a
Hermitian operator:
\beq
\hat{H}_{p}(t)=L(H_{pt}),\ \ \limht(\hat{H}_{p}(t))=H_{pt}.
\eeq
The lifting of $\tau$ is then defined by evolution under the Heisenberg
equations of motion. (Here Assumption~\ref{Ahc} is used.)  That
is $\tau$ is lifted to the unitary transformation defined by conjugation with
the unitary operator
\beq
\hat{U}_{p}=Te^{\frac{1}{i\hbar}\int_{0}^{1}\hat{H}_{p}(t')dt'},
\eeq
where $T$ denotes the time ordered product.

In general this construction is not enough to define a section.  However,
it is enough if to each point $\tau$ in $\Gamma$ we can associate a particular
path in $U$ connecting $T_{\Gamma}(\tau)$ to $T_{\Gamma}(\tau_{0})=I$, such
that the paths vary continuously
with $\tau$.  In this case we obtain a
section by assigning to each point $\tau$ in $\Gamma$ the unitary
transformation corresponding to the particular path $p_{\tau}$ which ends
at $T_{\Gamma}(\tau)$  (Presumably $p_{\tau_{0}}$ will be the constant
path so that $\tau_{0}$ lifts to the identity.):
\begin{eqnarray}
s:\Gamma&\rightarrow&\{(\tau,\hat{u})\in\Gamma\times\hat{U}\ |\ T_{\Gamma}
(\tau)=\rho(\hat{u})\}\nonumber\\
\tau&\mapsto&(\tau,\hat{u}_{\tau}),\nonumber\\
\hat{u}_{\tau}&=&unitary\ transformation\ implemented\ by\ conjugation\ with\
\nonumber\\
\hat{U}_{p_{\tau}}&=&Te^{\frac{1}{i\hbar}\int_{0}^{1}\hat{H}_{p_{\tau}}
(t')dt'}.
\label{n1}
\end{eqnarray}
Because the paths vary continuously with $\tau$,
$\hat{H}_{p_{\tau}}(t)$ varies continuously with $\tau$, and therefore
$\hat{U}_{p_{\tau}}$ varies continuously with $\tau$. Hence \rref{n1}
defines a section.
The condition that to each point $\tau$ in the base space $\Gamma$, we can
associate a particular $C^{1}$ path in $U$ connecting $T_{\Gamma}(\tau)$ to
$I$ , such that the
paths vary continuously with $\tau$ is Condition~\ref{Eco}.

We now make the assumption that the existence of a $C^{0}$ section implies
the existence of a $C^{1}$ section.  That is we assume some reasonable
compatibility between the topology and differential structure of the space of
operators on Hilbert space.  Then, if the provisions of Claim~\ref{Th1}
hold, a quantization after
evolution can be
carried out for a theory with global time $\Gamma$.  In the
next section we will show that Condition~\ref{Eco} holds when $\Gamma$ is a
subset of the space of slices parametrized by a CW-complex, and the spacetime
manifold is of the form $M=\Sigma\times\IR^{+}$.
The result will be interpreted in Section~\ref{Int2} and extended in
Section~\ref{qpt}.

\section{Consistent Evolution on a CW-Complex of Global Times}
\label{Con}

To obtain definite results, consideration will now be limited to a
subspace of the space of slices parametrized by a CW-complex.
Consider a spacetime manifold of the form $M=\Sigma\times\IR^{+}$, and take
$\sigma_{0}=\Sigma\times 0$.  Let $T_{2}$ be the mapping of the space of
slices $S$ into $U$ determined as in Section~\ref{Qua2} with $\sigma_{0}$
mapped to the identity.  Let $T_{1}$ be a mapping of a
CW-complex $\Gamma_{C}$ to $S$, which is a homeomorphism to its image,
such that there is a point $\tau_{0}$ which
maps to $\sigma_{0}$.  The diagram,
$$\begin{array}{ccccl}
\ &\ &\ &\ &\hat{U}\\
\ &\ &\ &\ &\downarrow \rho\\
\Gamma_{C}&\rightarrow&S&\rightarrow&U\ \ ,\\
\ &T_{1}&\ &T_{2}&\ \
\end{array}$$
leads to the pullback bundle over $\Gamma_{C}$,
$(P_{C},\Gamma_{C},\lambda_{C})$, defined by
\begin{eqnarray}
&P_{C}&=\{(\gamma,\hat{u})\in \Gamma_{C}\times\hat{U}\ |\ T_{2}\circ
T_{1}(\Gamma_{C})=\rho(\hat{u})\}\nonumber\\
&\lambda_{C}&:P_{C}\rightarrow \Gamma_{C}\nonumber\\
&(&\gamma,\hat{u})\mapsto \gamma.
\label{m1}
\end{eqnarray}
Of course the bundle \rref{m1} is the bundle \rref{m} with $\Gamma$ set
equal to $\Gamma_{C}$, a subspace of $S$ parametrized by a CW-complex.
A section of \rref{m1} provides a quantization after
evolution on the space of global time $\Gamma_{C}$.
We will show that Condition~\ref{Eco} (Section~\ref{Gen}) holds for
\rref{m1}.  Therefore, by
Claim~\ref{Th1},
a section of \rref{m1} exists if Assumption~\ref{Lim}, etc., is correct.

\begin{theo}
\label{Th2}
Let $E$ be the space of pairs of points $\tau$ in the CW-complex
$\Gamma_{C}$, and paths in $U$
connecting $T_{2}\circ T_{1}(\tau)$ to $T_{2}\circ T_{1}(\tau_{0})=I$:
$$E=\{(\tau,p)\ |\ \tau\in \Gamma_{C},\ p\ a\ C^{1}\ path\ in\ U\ connecting\
T_{2}\circ T_{1}(\tau)\ to\ I\}.$$
Then, if Assumption~\ref{Gti} (Section~\ref{Gra}) holds, the bundle
$(E,\Gamma_{C},\sigma)$
defined by
\begin{eqnarray*}
\sigma:E&\rightarrow& \Gamma_{C},\\
(\tau,p)&\mapsto& \tau,
\end{eqnarray*}
has a  section.
\end{theo}

To argue for Claim~\ref{Th2} consider the smaller bundle obtained by
restricting
$E$ to paths that are the images of time-slicings (A time-slicing is a path
in $S$, but not all paths in $S$ are time-slicings.):
\begin{eqnarray}
&E&^{\prime}=\{(\tau,p)\ |\ \tau\in \Gamma_{C},\ p\ a\ time\ slicing\
connecting\ T_{1}(\tau)\
to\ T_{1}(\tau_{0})\},\nonumber\\
&\sigma&^{\prime}:E^{\prime}\rightarrow \Gamma_{C},\nonumber\\
&(&\tau,p)\mapsto \tau.
\label{epr}
\end{eqnarray}
Of course if $(E^{\prime},\Gamma_{C},\sigma^{\prime})$ has a section then so
does
$(E,\Gamma_{C},\sigma)$.  We will use a theorem due to Gromov
\cite{Gro} to show that
the fiber $\sigma^{\prime -1}(\gamma)$, $\gamma\in \Gamma_{C}$, is weak
homotopy equivalent to a point.  In particular, all the homotopy groups
are trivial.  Using arguments from obstruction theory \cite{Span}, this fact,
together with the fact that $\Gamma_{C}$ is a CW-complex, can be used to show
that a section exists.

Gromov's theorem is quite general.  The basic result we will need is that
the space of spacelike $(n-1)$-plane fields on an n-dimensional non-closed
Lorentzian manifold is weak homotopy equivalent to the space of integrable
spacelike $(n-1)$-plane fields.  A $p$-plane field is an assignment of a
$p$-dimensional subspace of the tangent space to each point in a manifold.
It can also be described as a section in the fiber bundle associated with
the tangent bundle, with fibers the space of $p$-dimensional subspaces of
the tangent space.  An integrable spacelike $(n-1)$-dimensional plane
field on an $n$-dimensional Lorentzian manifold is a time-slicing\footnote{
Actually it is a foliation with spacelike leaves.  By Assumption~\ref{Gti}
there is a homeomorphism between the space of such foliations and the space
of time-slicings.  Here we fix a spacetime and use this homeomorphism to
to obtain topological statements about time-slicings.}.

\begin{picture}(300,220)(-45,-50)
\newsavebox{\side}
\newsavebox{\elipse}
\savebox{\hor}(0,0)[l]{\line(1,0){250}}
\savebox{\side}(0,0)[bl]{\line(1,1){100}}
\savebox{\elipse}(0,0){
\qbezier(50,-12.5)(0,0)(50,12.5)
\qbezier(50,12.5)(100,25)(150,12.5)
\qbezier(150,12.5)(200,0)(150,-12.5)
\qbezier(150,-12.5)(100,-25)(50,-12.5)}
\put(72.5,52.5){\usebox{\elipse}}
\put(0,0){\usebox{\hor}}
\put(100,100){\line(1,0){233.3333}}
\put(0,0){\usebox{\side}}
\put(250,0){\line(5,6){83.3333}}
\put(170,52.5){\line(1,1){100}}
\put(170,52.5){\line(-1,1){100}}
\put(170,52.5){\vector(0,1){60}}
\put(170,52.5){\vector(2,3){40}}
\multiput(172.5,112.5)(10,0){4}{\line(1,0){5}}
\put(170,52.5){\vector(1,0){40}}
\multiput(210,55)(0,10){6}{\line(0,1){5}}
\put(156,85){$\frac{\partial}{\partial t}$}
\put(185,41.5){$\vec{x}$}
\put(153,48.25){\vector(4,1){17}}
\put(140,45){\vector(-4,-1){18}}
\put(144,43){$1$}
\put(20,-30){Figure 5\ \ \ \ The fibers of $(Y,X,\alpha)$ are convex domains}
\sbox{\hor}{}
\sbox{\side}{}
\sbox{\elipse}{}
\end{picture}

If we take as the non-closed manifold the piece of the spacetime manifold
between $T_{1}(\tau_{0})$ and $T_{1}(\tau)$ together with $T_{1}(\tau_{0})$
and $T_{1}(\tau)$ (a manifold
with boundary), call it $X$, then a time-slicing necessarily includes
$T_{1}(\tau_{0})$ and $T_{1}(\tau)$ as slices.  The space of time-slicings
of $X$ is the fiber of $(E^{\prime},\Gamma_{C},\sigma^{\prime})$ over $\tau$.
Therefore,
if we can show that the space of spacelike $(n-1)$-plane fields on $X$ is
contractible, then by Gromov's theorem the homotopy groups of the fibers
of $(E^{\prime},\Gamma_{C},\sigma^{\prime})$ are all trivial.

Let $(Y,X,\alpha)$ be the fiber bundle associated with the tangent
bundle of $X$ with fibers the space of spacelike $(n-1)$-dimensional
subspaces
of the tangent space.  As stated earlier, the space of sections of
$(Y,X,\alpha)$ is the space of spacelike $(n-1)$-plane fields.
First we will show that if the fibers of $(Y,X,\alpha)$ are convex
domains, then the space of sections of $(Y,X,\alpha)$ is a convex
domain.  Let $s_{1}:X\rightarrow Y$ and $s_{2}:X\rightarrow Y$ be two sections
of $(Y,X,\alpha)$.  Then for $x$ in $X$,  $s_{1}(x)$ and
$s_{2}(x)$ are in $\alpha^{-1}(x)$, and
$$(\lambda s_{1}+(1-\lambda) s_{2})(x)=\lambda s_{1}(x)+(1-\lambda) s_{2}(x)$$
is in $\alpha^{-1}(x)$ if $\alpha^{-1}(x)$ is convex.  Therefore
$\lambda s_{1}+(1-\lambda) s_{2}$ is a section and the space of sections is
convex if the fibers of $(Y,X,\alpha)$ are convex.

We must show that each fiber of $(Y,X,\alpha)$---that is the space
of $(n-1)$-dimensional spacelike subspaces of the tangent space---is a convex
domain.  Each tangent space of $X$ acquires a Minkowski metric from the
Lorentzian metric on $X$.  By choosing a coordinate system in which the
metric is diag$(-1,1,1,\ldots)$, the spacelike subspaces are put in one to
one correspondence with the time-like vectors of the form $(1,\vec{x})$
(see Figure 5).
The time-like vectors of the form $(1,\vec{x})$ are in one to one
correspondence
with the vectors in $\IR^{n-1}$ of length less then one.  (The former are
a graph over the latter.)  The space of vectors in $\IR^{n-1}$ of length less
then one is clearly a convex domain. Hence each fiber of
$(Y,X,\alpha)$ acquires the structure of a convex domain.

\begin{picture}(300,180)(-70,-60)
\newsavebox{\lside}
\newsavebox{\corner}
\newsavebox{\arrow}
\savebox{\corner}(0,0)[l]{\circle*{4}}

\savebox{\hor}(0,0)[l]{\thicklines{\line(1,0){100}}}

\put(0,0){\usebox{\hor}}
\put(33,50){\thicklines{\line(1,0){91.6667}}}
\savebox{\side}(0,0)[bl]{\thicklines{\line(2,3){33}}}
\put(0,0){\usebox{\side}}
\put(100,0){\thicklines{\line(1,2){25}}}
\put(0,0){\circle*{4}}
\put(100,0){\circle*{4}}
\put(33,50){\circle*{4}}
\put(124.66667,50){\circle*{4}}

\savebox{\lside}(0,0)[b]{\thicklines{\line(0,1){40}}}

\put(0,40){\usebox{\lside}}
\put(100,40){\usebox{\lside}}
\put(0,70){\circle*{4}}
\put(100,55){\circle*{4}}

\savebox{\arrow}(0,0)[b]{\thicklines{\vector(0,1){20}}}

\put(0,10){\usebox{\arrow}}
\put(100,10){\usebox{\arrow}}

\put(200,0){\usebox{\hor}}
\put(200,40){\usebox{\hor}}
\put(200,80){\usebox{\hor}}
\put(200,40){\usebox{\lside}}
\put(300,40){\usebox{\lside}}
\put(250,10){\usebox{\arrow}}
\thicklines{\qbezier(200,70)(215,55)(230,60)}
\thicklines{\qbezier(230,60)(260,70)(300,55)}
\put(200,0){\circle*{4}}
\put(300,0){\circle*{4}}
\put(200,70){\circle*{4}}
\put(300,55){\circle*{4}}
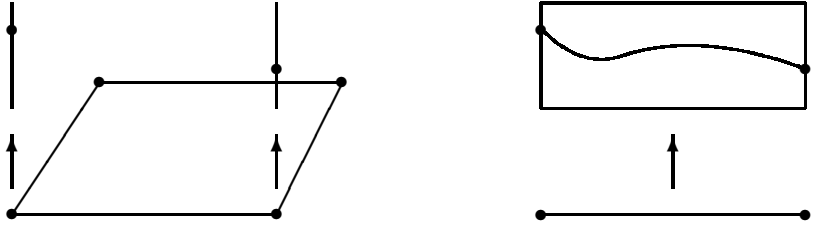
\put(-10,-40){Figure 6\ \ \ \ Construction of a section over a CW-complex}
\sbox{\hor}{}
\sbox{\side}{}
\sbox{\lside}{}
\sbox{\corner}{}
\sbox{\arrow}{}
\end{picture}

Following the above line of reasoning, we conclude that the homotopy groups of
the fibers of $(E^{\prime},\Gamma_{C},\sigma^{\prime})$ are all trivial.
This means
that for
all $m$, any mapping of the sphere $S^{m}$ into a fiber of
$(E^{\prime},\Gamma_{C},\sigma^{\prime})$ is extendible to a mapping of the
ball
$D^{m+1}$
into the fiber.  Begin constructing a section of
$(E^{\prime},\Gamma_{C},\sigma^{\prime})$ by
choosing arbitrary liftings of the zero cells of $\Gamma_{C}$ to $E^{\prime}$
(see Figure 6).
The pullback bundle over a single cell is necessarily a trivial fiber bundle,
because a cell is contractible and all the fibers are homeomorphic.  Therefore,
because any mapping of $S^{0}$ (two points) to a fiber may be extended to a
mapping of $D^{1}$ to the fiber, the lifting of the zero cells may be extended
to a lifting of the 1-cells.  The argument repeats for the 2-cells, 3-cells,
$\ldots$ , on up to $\infty$.  There is not even a requirement that the number
of cells be countable.

We have shown that a section of $(E^{\prime},\Gamma_{C},\sigma^{\prime})$ ,
and
hence of
$(E,\Gamma_{C},\sigma)$, exists for an arbitrary CW-complex
$\Gamma_{C}$.
Therefore, Condition~\ref{Eco} is satisfied.  Provided Assumption~\ref{Lim}
holds, etc., we may have
consistent evolution on any subspace of the space of global time
parametrized by a (finite or infinite) CW-complex.

\section{Interpretation of a CW-Complex of Global Times}
\label{Int2}

How much of $S$ may be parametrized by a CW-complex?  That is, how much
does this result say about the general problem of consistent evolution?  Let
us consider the question:  given a particular time-slicing connecting
$\sigma_{0}$ to $\sigma$, how many other time-slicings
connecting $\sigma_{0}$ and $\sigma$ can define the same evolution?

A particular time-slicing connecting $\sigma_{0}$ to $\sigma$ is a family of
slices (global times) parametrized by the closed interval $[0,1]$.
A family of deformations of the time-slicing, parametrized by another closed
interval, is a family of slices parametrized by a closed disc.  A family of
deformations of the time-slicing parametrized by $[0,1]\times [0,1]$, is a
family of slices parametrized by a closed 3-ball, etc.  Clearly we may have
an infinite parameter family of deformations of the time-slicing with slices
parametrized by a CW-complex.  If there is any particular time-slicing we
wish to include, we may always include it\footnote{The space of time-slicings
is connected, because its $0$-th
homotopy group is trivial.  Therefore we may reach any time-slicing by
deforming the original time-slicing.}.  By Claims \ref{Th1} and \ref{Th2} all
these time-slicings may be brought into agreement (if Assumption~\ref{Lim},
etc., holds).  Furthermore, if the time-slicings include common intermediate
slices, they may be brought into agreement on those slices as well.

The situation is probably something like
trying to cover the real number line with a closed interval.  You can always
add in any additional points you like. But you can never cover the whole
real number line.
If we could show that the space of slices $S$ was
homotopic to a CW-complex, then we could show that Condition~\ref{Eco}
(Section~\ref{Gen}) holds
for $S$.  There is a theorem due to John Milnor \cite{Mil} which states
roughly:
``The space of mappings of a compact topological space to a CW-complex
is homotopic to a CW-complex.''  Well, the space $S$ is the
space of spacelike mappings of $\Sigma$ to $M=\Sigma\times \IR^{+}$.  It
seems likely that a proof of the general result may proceed along these lines.

\section{Quantization by Preferred Time-Slicing}
\label{qpt}

Arguably the most practical approach to making a physical prediction using
quantum gravity
is to reduce gravity to a Hamiltonian system using a convenient time-slicing
connecting the initial and final slices.  Assuming that Assumption~\ref{Ach}
(Section~\ref{Com})
is correct, this reduces the problem to quantizing a Hamiltonian system with
the attendant operator ordering ambiguities.  Let us avoid the many conceptual
issues by assuming that this at least would be sensible and correct if we
used the right time-slicing and the right operator ordering.  Then assuming
that physics requires consistent evolution with respect to all time-slicings,
there must be an operator ordering which is right for our ``most convenient''
time-slicing.  Lacking experimental data, we will have to search for this
ordering by invoking some esthetic principle or new physical theory.  An
interesting question is: is there any ordering choice which would preclude
the possibility of consistent evolution with respect to some other
time-slicing, and hence be excluded on this basis?  At least for an infinite
parameter family of time-slicings---more precisely a subspace of $S$
parametrized by a CW-complex---if the
assumptions of this paper are correct the answer is no.

To see this let $\Gamma_{C}$ be as in Section~\ref{Con}, and let the initial
and final slices be $\sigma_{i}$ and $\sigma_{f}$ respectively (they
should be in the image of $T_{1}$).  If $\sigma_{i}$ and $\sigma_{f}$ do
not correspond to
zero cells, further subdivide $\Gamma_{C}$ so that they do.  There exists a
time-slicing $T$ containing $\sigma_{0},\ \sigma_{i}$, and $\sigma_{f}$
which coincides with the ``most convenient'' time-slicing between $\sigma_{i}$
and $\sigma_{f}$.  Now construct a quantization after evolution of the
system with global time $\Gamma_{C}$ by choosing a section of \rref{epr}
and a lifting $L$ of the real valued functions on phase space to
Hermitian operators.  Following the argument for Claim~\ref{Th2}
(Section~\ref{Con}) we find
that the liftings of the zero cells of $\Gamma_{C}$ are arbitrary.  Hence in
choosing the section of \rref{epr}, we can lift the zero cell corresponding
to $\sigma_{f}$ to $T$, and the zero cell corresponding to $\sigma_{i}$ to
the portion of $T$ between $\sigma_{0}$ and $\sigma_{i}$.
If $H_{T}(t)$ ($t\in[0,1]$) is the classical
Hamiltonian for the time-slicing $T$, then we now have a quantization after
evolution where evolution from $\sigma_{i}$ to $\sigma_{f}$ is
consistent with evolution along the most convenient time-slicing with the
operator Hamiltonian $L(H_{T}(t))$.  Given any operator valued continuous
function $\hat{f}$ on $S$, with $\limht(\hat{f}(\sigma))=0$ for all $\sigma$,
we can always redefine $L$
by\footnote{If $\hat{f}$ is not constant then $L$ will become
slice-dependent; but this causes no problem.}
$$L(H(t))\rightarrow L^{\prime}(H(t))=L(H(t))+\hat{f}(\sigma(t)).$$
Hence, $L(H_{T}(t))$ can have any operator ordering whatsoever.

\section{Conclusion}
\label{Con2}

The relation of this work to gravity is based on Assumption~\ref{Gti}
(Section~\ref{Gra}),
essentially that gravity may be reduced to an unconstrained system with the
``space of slices'' playing the role of time.  If this assumption fails, then
this work may be applicable to some other theory, but not gravity.
If Assumption~\ref{Lim} (Section~\ref{Cla}) fails, that is if it is not
locally possible to
continuously
lift the real valued functions on phase space to Hermitian operators, then
the result of this work is only to place further doubt on the possibility for
consistent evolution.
There are also a few seemingly innocuous assumptions: paracompactness,
deformability of continuous mappings to smooth ones, and
Assumption~\ref{Ahc} (Section~\ref{Com}).
To avoid repeating them over and over again, let us
assume that all the above assumptions are correct for the remainder of this
discussion.

This said, to interpret our results, it must be known if
quantization after evolution really is equivalent to Heisenberg equation
evolution.  (Here we mean Assumption~\ref{Ach} in Section~\ref{Com}.)
If it is, then the
interpretation is that it is possible
to operator order the myriad of Hamiltonians, corresponding to an infinite
parameter family of time-slicings (at least), so that evolution is consistent.
However, time dependent operator orderings of the Hamiltonians may be required.
In addition any particular time-slicing, with evolution
defined by any operator ordering of the associated Hamiltonian, may be
included in such a consistent infinite parameter family.

If, however, quantization after evolution is not equivalent to Heisenberg
equation evolution, that is if there is some exotic form of differentiable
unitary evolution
with the correct classical limit, but which is generated by a Hamiltonian with
the wrong or possibly no classical limit, then the interpretation depends on
whether the Hamiltonians should be considered observables.  This, I suppose,
depends on whether or not the Hamiltonians can be measured.  Observables
which are defined on slices are fine.  They evolve consistently and have the
right classical limit.  But the Hamiltonians are defined not on slices, but
on local time-slicings.  If this makes them unmeasurable, then quantization
after evolution is still a reasonable physical theory.  However, if
Hamiltonians in geometrodynamics are measurable, then quantization after
evolution must be equivalent to Heisenberg equation evolution if it is to have
the right classical limit.

Of course one should at least ask whether or not evolution along
different time-slicings {\em should} agree.  In the past we have found that the
existence of a classical symmetry, that is the fact that the classical theory
transformed under a trivial representation of some transformation group, did
{\em not} always mean that the quantum theory transformed under a trivial
representation.  The classic example is the electron and its transformation
under rotations.  Electron spin is a purely quantum mechanical effect and
possibly some phenomena involving different time-slicings also exists.
However, this would mean that the unified picture of space and time, so
treasured by relativists, would not hold in the quantum theory.
What this work has attempted to show is that such a result is not forced on us.

\begin{flushleft}
{\large\bf Acknowledgments}
\end{flushleft}
Steve Carlip provided valuable criticisms throughout this work.  Dmitry
Fuchs suggested the proof of Claim~\ref{Th2} as well as providing
other essential help
with topology.  I would like to thank them both.  This work was supported
in part by National Science Foundation grant NSF-PHY-93-57203 and Department
of Energy grant DE-FG03-91ER40674.



\end{document}